\documentclass[a4paper,12pt]{iopart}
\usepackage{iopams}
\usepackage{setstack}
\usepackage{graphicx}
\usepackage{bm}
\usepackage{epsfig}
\newcommand{\diag}{{\rm diag\,}}
\newcommand{\Str}{{\rm Str\,}}
\newcommand{\Sdet}{{\rm Sdet\,}}
\newcommand{\UOSp}{{\rm UOSp\,}}
\newcommand{\USp}{{\rm USp\,}}
\newcommand{\U}{{\rm U\,}}
\newcommand{\CU}{{\rm CU\,}}
\newcommand{\Herm}{{\rm Herm\,}}
\newcommand{\RE}{{\rm Re\,}}
\newcommand{\IM}{{\rm Im\,}}
\newcommand{\eins}{\leavevmode\hbox{\small1\kern-3.8pt\normalsize1}}
\eqnobysec

\begin{document}

\newtheorem{definition}{Definition}[section]
\newtheorem{assumption}[definition]{Assumption}
\newtheorem{theorem}[definition]{Theorem}
\newtheorem{lemma}[definition]{Lemma}
\newtheorem{corollary}[definition]{Corollary}

\title[Superbosonization formula and generalized Hubbard--Stratonovich transformation]{Comparison of the superbosonization formula and the generalized Hubbard--Stratonovich transformation}
\author{Mario Kieburg$^\dagger$, Hans-J\"urgen Sommers and Thomas Guhr}
\address{Universit\"at Duisburg-Essen, Lotharstra\ss e 1, 47048 Duisburg, Germany}
\eads{$^\dagger$ \mailto{mario.kieburg@uni-due.de}}

\date{\today}

\begin{abstract}
Recently, two different approaches were put forward to extend the supersymmetry method in random matrix theory from Gaussian ensembles to general rotation invariant ensembles. These approaches are the generalized Hubbard--Stratonovich transformation and the superbosonization formula. Here, we prove the equivalence of both approaches. To this end, we reduce integrals over functions of supersymmetric Wishart--matrices to integrals over quadratic supermatrices of certain symmetries.
\end{abstract}

\submitto{\JPA}
\maketitle

\section{Introduction}\label{sec1}

The supersymmetry technique is a powerful method in random matrix theory and disordered systems. For a long time it was thought to be applicable for Gaussian probability densities only \cite{Efe83,VerZir85,VWZ85,Efe97}. Due to universality on the local scale of the mean level spacing \cite{BreZee93a,BreZee93b,HacWei95,GMW98}, this restriction was not a limitation for calculating in quantum chaos and disordered systems. Indeed, results of Gaussian ensembles are identical for large matrix dimension with other invariant matrix ensembles on this scale. In the Wigner--Dyson theory \cite{Bee97} and its corrections for systems with diffusive dynamics \cite{Mir00}, Gaussian ensembles are sufficient. Furthermore, universality was found on large scale, too \cite{AJM90}. This is of paramount importance when investigating matrix models in high--energy physics.

There are, however, situations in which one can not simply resort to Gaussian random matrix ensembles. The level densities in high--energy physics \cite{BIPZ78} and finance \cite{LCBP99} are needed for non-Gaussian ensembles. But these one--point functions strongly depend on the matrix ensemble. Other examples are bound--trace and fixed--trace ensembles \cite{Meh04}, which are both norm--dependent ensembles \cite{Guh06}, as well as ensembles derived from a non-extensive entropy principle \cite{TVT04,BBP04,Abu04}. In all these cases one is interested in the non-universal behavior on special scales.

Recently, the supersymmetry method was extended to general rotation invariant probability densities \cite{Guh06,Som07,LSZ07,KGG08}. There are two approaches. The first one is the generalized Hubbard--Stratonovich transformation \cite{Guh06,KGG08}. With help of a proper Dirac--distribution in superspace an integral over rectangular supermatrices was mapped to a supermatrix integral with non-compact domain in the Fermion--Fermion block. The second approach is the superbosonization formula \cite{Som07,LSZ07} mapping the same integral over rectangular matrices as before to a supermatrix integral with compact domain in the Fermion--Fermion block.

In this work, we prove the equivalence of the generalized Hubbard--Stratonovich transformation with the superbosonization formula. The proof is based on integral identities between supersymmetric Wishart--matrices and quadratic supermatrices. The orthogonal, unitary and unitary-symplectic classes are dealt with in a unifying way.

The article is organized as follows. In Sec. \ref{sec2}, we give a motivation and introduce our notation. In Sec. \ref{sec3}, we define rectangular supermatrices and the supersymmetric version of Wishart-matrices built up by supervectors. We also give a helpful corollary for the case of arbitrary matrix dimension discussed in Sec. \ref{sec7}. In Secs. \ref{sec4} and \ref{sec5}, we present and further generalize the superbosonization formula and the generalized Hubbard--Stratonovich transformation, respectively. The theorem stating the equivalence of both approaches is given in Sec. \ref{sec6} including a clarification of their mutual connection. In Sec. \ref{sec7}, we extend both theorems given in Secs. \ref{sec4} and \ref{sec5} to arbitrary matrix dimension. Details of the proofs are given in the appendices.

\section{Ratios of characteristic polynomials}\label{sec2}

We employ the notation defined in Refs.  \cite{KKG08,KGG08}. $\Herm(\beta,N)$ is either the set of $N\times N$ real symmetric ($\beta=1$), $N\times N$ hermitian ($\beta=2$) or $2N\times 2N$ self-dual ($\beta=4$) matrices, according to the Dyson--index $\beta$. We use the complex representation of the quaternionic numbers $\mathbb{H}$. Also, we define
\begin{equation}\label{2.0}
 \gamma_1=\left\{\begin{array}{ll} 1 & ,\ \beta\in\{2,4\} \\ 2 & ,\ \beta=1 \end{array}\right. \quad ,\quad \gamma_2=\left\{\begin{array}{ll} 1 & ,\ \beta\in\{1,2\} \\ 2 & ,\ \beta=4 \end{array}\right.
\end{equation}
and $\tilde{\gamma}=\gamma_1\gamma_2$.

The central objects in many applications of supersymmetry are averages over ratios of characteristic polynomials \cite{AkeFyo03,AkePot04,BorStr05}
\begin{eqnarray}
  \fl Z_{k_1k_2}(E^-) & = & \int\limits_{\Herm(\beta,N)}P(H)\frac{\prod\limits_{n=1}^{k_2}\det\left(H-(E_{n2}-\imath\varepsilon)\eins_{\gamma_2N}\right)}{\prod\limits_{n=1}^{k_1}\det\left(H-(E_{n1}-\imath\varepsilon)\eins_{\gamma_2N}\right)}d[H]\nonumber\\
  & = & \int\limits_{\Herm(\beta,N)}P(H)\Sdet^{-1/\tilde{\gamma}}\left(H\otimes \eins_{\tilde{\gamma}(k_1+k_2)}- \eins_{\gamma_2N}\otimes E^-\right)d[H]\label{2.1}
\end{eqnarray}
where $P$ is a sufficiently integrable probability density on the matrix set $\Herm(\beta,N)$ invariant under the group
\begin{equation}\label{2.2}
 \U^{(\beta)}(N)=\left\{\begin{array}{ll} 
              \Or(N)	& ,\ \beta=1\\
	      \U(N)	& ,\ \beta=2\\
	      \USp(2N)	& ,\ \beta=4
             \end{array}\right. .
\end{equation}
Here, we assume that $P$ is analytic in its real independent variables. We use the same measure for $d[H]$ as in Ref.  \cite{KKG08} which is the product over all real independent differentials, see also Eq.~\eref{t1.4}. Also, we define $E=\diag(E_{11},\ldots,E_{k_11},E_{12},\ldots,E_{k_22})\otimes \eins_{\tilde{\gamma}}$ and $E^-=E-\imath\varepsilon \eins_{\tilde{\gamma}(k_1+k_2)}$.

The generating function of the $k$--point correlation function \cite{BreHik00,Zir06,Guh06,KGG08}
\begin{equation}\label{2.3}
 R_{k}(x)=\gamma_2^{-k}\int\limits_{\Herm(\beta,N)}P(H)\prod\limits_{p=1}^k\tr\delta(x_p-H)d[H]
\end{equation}
is one application and can be computed starting from the matrix Green function and Eq.~\eref{2.1} with $k_1=k_2=k$. Another example is the $n$--th moment of the characteristic polynomial \cite{MehNor01,Fyo02,Zir06}
\begin{equation}\label{2.4}
 \widehat{Z}_{n}(x,\mu)=\int\limits_{\Herm(\beta,N)}P(H)\Theta(H){\det}^n\left(H-E \eins_{\gamma_2k}\right)d[H],
\end{equation}
where the Heavyside--function for matrices $\Theta(H)$ is unity if $H$ is positive definite and zero otherwise. \cite{KGG08} 

With help of Gaussian integrals, we get an integral expression for the determinants in Eq.~\eref{2.1}. Let $\Lambda_j$ be the Grassmann space of $j$--forms. We consider a complex Grassmann algebra \cite{Ber87} $\Lambda=\bigoplus\limits_{j=0}^{2\gamma_2Nk_2}\Lambda_j$ with $\gamma_2Nk_2$ pairs $\{\zeta_{jn},\zeta_{jn}^*\}$, $1\leq n\leq k_2,\ 1\leq j\leq \gamma_2N$, of Grassmann variables and use the conventions of Ref.  \cite{KKG08} for integrations over Grassmann variables. Due to the $\mathbb{Z}_2$--grading, $\Lambda$ is a direct sum of the set of commuting variables $\Lambda^0$ and of anticommuting variables $\Lambda^1$. The body of an element in $\Lambda$ lies in $\Lambda_0$ while the Grassmann generators are elements in $\Lambda_1$.

Let $\imath$ be the imaginary unit. We take $\gamma_2Nk_1$ pairs $\{z_{jn},z_{jn}^*\}$, $1\leq n\leq k_1,\ 1\leq j\leq \gamma_2N$, of complex numbers and find for Eq.~\eref{2.1}
\begin{equation}\label{2.5}
 \fl Z_{k_1k_2}(E^-)= (2\pi)^{\gamma_2N(k_2-k_1)}\imath^{\gamma_2Nk_2}\int\limits_{\mathfrak{C}} \mathcal{F}P(K)\exp\left(-\imath\Str BE^-\right)d[\zeta]d[z]
\end{equation}
where $d[z]=\prod\limits_{p=1}^{k_1}\prod\limits_{j=1}^{\gamma_2N}dz_{jp}dz_{jp}^*$ , $d[\zeta]=\prod\limits_{p=1}^{k_2}\prod\limits_{j=1}^{\gamma_2N}(d\zeta_{jp}d\zeta_{jp}^*)$ and $\mathfrak{C}=\mathbb{C}^{\gamma_2k_1N}\times\Lambda_{2\gamma_2Nk_2}$. The characteristic function appearing in \eref{2.5} is defined as
\begin{equation}\label{2.6}
 \mathcal{F}P(K)=\int\limits_{\Herm(\beta,N)}P(H)\exp\left(\imath\tr HK\right)d[H] .
\end{equation}
The two matrices
\begin{equation}\label{2.7}
 K = \frac{1}{\tilde{\gamma}}V^\dagger V\qquad{\rm and}\qquad
 B = \frac{1}{\tilde{\gamma}}VV^\dagger
\end{equation}
are crucial for the duality between ordinary and superspace. While $K$ is a $\gamma_2N\times\gamma_2N$ ordinary matrix whose entries have nilpotent parts, $B$ is a $\tilde{\gamma}(k_1+k_2)\times\tilde{\gamma}(k_1+k_2)$ supermatrix. They are composed of the rectangular $\gamma_2N\times\tilde{\gamma}(k_1+k_2)$ supermatrix
\begin{eqnarray}
 V^\dagger|_{\beta\neq2} & = & (z_1,\ldots,z_{k_1},Yz_1^*,\ldots,Yz_{k_1}^*,\zeta_1,\ldots,\zeta_{k_2},Y\zeta_1^*,\ldots,Y\zeta_{k_2}^*) ,\nonumber\\
 V|_{\beta\neq2} & = & (z_1^*,\ldots,z_{k_1}^*,Yz_1,\ldots,Yz_{k_1},-\zeta_1^*,\ldots,-\zeta_{k_2}^*,Y\zeta_1,\ldots,Y\zeta_{k_2})^T ,\nonumber\\
 V^\dagger|_{\beta=2} & = & (z_1,\ldots,z_{k_1},\zeta_1,\ldots,\zeta_{k_2}) ,\nonumber\\
 V|_{\beta=2} & = & (z_1^*,\ldots,z_{k_1}^*,-\zeta_1^*,\ldots,-\zeta_{k_2}^*)^T .\label{2.9}
\end{eqnarray}
The transposition ``$T$'' is the ordinary transposition and is not the supersymmetric one. However, the adjoint ``$\dagger$'' is the complex conjugation with the supersymmetric transposition ``$T_{\rm S}$''
\begin{equation}\label{2.9b}
 \sigma^{T_{\rm S}}=\left[\begin{array}{cc} \sigma_{11} & \sigma_{12} \\ \sigma_{21} & \sigma_{22} \end{array}\right]^{T_{\rm S}}=\left[\begin{array}{cc} \sigma_{11}^T & \sigma_{21}^T \\ -\sigma_{12}^T & \sigma_{22}^T \end{array}\right],
\end{equation}
where $\sigma$ is an arbitrary rectangular supermatrix. We introduce the constant $\gamma_2N\times\gamma_2N$ matrix
\begin{equation}\label{2.10}
 Y=\left\{\begin{array}{ll}
             \eins_N & ,\ \beta=1\\
	     Y_s^T\otimes \eins_N & ,\ \beta=4
   \end{array}\right.\ ,\ \ Y_s=\left[\begin{array}{cc} 0 & 1 \\ -1 & 0 \end{array}\right] .
\end{equation}
The crucial duality relation \cite{Guh06,KGG08}
\begin{equation}\label{2.11}
 \tr K^m=\Str B^m\ ,\ \ m\in\mathbb{N} ,
\end{equation}
holds, connecting invariants in ordinary and superspace. As $\mathcal{F}P$ inherits the rotation invariance of $P$, the duality relation \eref{2.11} yields
\begin{equation}\label{2.12}
 \fl Z_{k_1k_2}(E^-)= (2\pi)^{\gamma_2N(k_2-k_1)}\imath^{\gamma_2Nk_2}\int\limits_{\mathfrak{C}} \Phi(B)\exp\left(-\imath\Str BE^-\right)d[\zeta]d[z] .
\end{equation}
Here, $\Phi$ is a supersymmetric extension of a representation $\mathcal{F}P_0$ of the characteristic function,
\begin{equation}\label{2.13}
 \Phi(B)=\mathcal{F}P_0(\Str B^m|m\in\mathbb{N})=\mathcal{F}P_0(\tr K^m|m\in\mathbb{N})=\mathcal{F}P(K) .
\end{equation}
The representation $\mathcal{F}P_0$ is not unique \cite{BEKYZ07}. However, the integral \eref{2.12} is independent of a particular choice \cite{KGG08}.

The supermatrix $B$ fulfills the symmetry
\begin{equation}\label{2.14}
 B^*=\left\{\begin{array}{ll}
                \widetilde{Y}B\widetilde{Y}^T &,\ \beta\in\{1,4\}, \\
		\widetilde{Y}B^*\widetilde{Y}^T &,\ \beta=2
               \end{array}\right.
\end{equation}
with the supermatrices
\begin{equation}\label{2.15}
 \fl\widetilde{Y}|_{\beta=1}=\left[\begin{array}{ccc} 0 &  \eins_{k_1} & 0 \\  \eins_{k_1} & 0 & 0 \\ 0 & 0 & Y_s\otimes \eins_{k_2} \end{array}\right]\quad,\quad\widetilde{Y}|_{\beta=4}=\left[\begin{array}{ccc} Y_s\otimes \eins_{k_1} & 0 & 0 \\ 0 & 0 &  \eins_{k_2} \\ 0 &  \eins_{k_2} & 0 \end{array}\right]
\end{equation}
and $\widetilde{Y}|_{\beta=2}=  \eins_{k_1+k_2}$ and is self-adjoint for every $\beta$. Using the $\pi/4$--rotations
\begin{equation}\label{2.16}
 \fl U|_{\beta=1}=\frac{1}{\sqrt{2}}\left[\begin{array}{ccc}  \eins_{k_1} &  \eins_{k_1} & 0 \\ -\imath \eins_{k_1} & \imath \eins_{k_1} & 0 \\ 0 & 0 & \sqrt{2}\  \eins_{2k_2} \end{array}\right],\ U|_{\beta=4}=\frac{1}{\sqrt{2}}\left[\begin{array}{ccc} \sqrt{2}\  \eins_{2k_1} & 0 & 0 \\ 0 &  \eins_{k_2} &  \eins_{k_2}\\ 0 & -\imath \eins_{k_2} & \imath \eins_{k_2} \end{array}\right]
\end{equation}
and $U|_{\beta=2}= \eins_{k_1+k_2}$, $\widehat{B}=UBU^\dagger$ lies in the well-known symmetric superspaces \cite{Zir96},
\begin{eqnarray}
 \fl\widetilde{\Sigma}_{\beta,\gamma_1k_1,\gamma_2k_2}^{(\dagger)} & = &\Biggl\{\sigma\in{\rm Mat}(\tilde{\gamma}k_1/\tilde{\gamma}k_2)\Biggl|\sigma^\dagger=\sigma,\nonumber\\
 & & \left.\sigma^*=\left\{\begin{array}[c]{ll}
                \widehat{Y}_{\gamma_1k_1,\gamma_2k_2}\sigma\widehat{Y}_{\gamma_1k_1,\gamma_2k_2}^{T} & ,\ \beta\in\{1,4\} \\
		\widehat{Y}_{k_1k_2}\sigma^*\widehat{Y}_{k_1k_2}^{T} & ,\ \beta=2
          \end{array}\right\}\right\}\label{2.17}
\end{eqnarray}
where
\begin{equation}\label{2.18}
 \fl\left.\widehat{Y}_{pq}\right|_{\beta=1}=\left[\begin{array}{cc}   \eins_{p} & 0 \\ 0 & Y_s\otimes \eins_{q} \end{array}\right]\ ,\ \ \left.\widehat{Y}_{pq}\right|_{\beta=2}=  \eins_{p+q}\ \ \ {\rm and}\ \ \ \left.\widehat{Y}_{pq}\right|_{\beta=4}=\left[\begin{array}{cc} Y_s\otimes \eins_{p} & 0 \\ 0 &  \eins_{q} \end{array}\right].
\end{equation}
The set ${\rm Mat}(p/q)$ is the set of $(p+q)\times(p+q)$ supermatrices on the complex Grassmann algebra $\bigoplus\limits_{j=0}^{2pq}\Lambda_j$. The entries of the diagonal blocks of an element in ${\rm Mat}(p/q)$ lie in $\Lambda^0$ whereas the entries of the off-diagonal block are elements in $\Lambda^1$.

The rectangular supermatrix $\widehat{V}^\dagger=V^\dagger U^\dagger$ is composed of real, complex or quaternionic supervectors whose adjoints form the rows. They are given by
\begin{equation}\label{2.19}
 \fl\Psi_j^\dagger=\left\{\begin{array}{ll}
                        \left(\left\{\sqrt{2}\RE z_{jn},\sqrt{2}\IM z_{jn}\right\}_{1\leq n\leq k_1},\left\{\zeta_{jn},\zeta^*_{jn}\right\}_{1\leq n\leq k_2}\right) & ,\ \beta=1,\\
                        \left(\left\{z_{jn}\right\}_{1\leq n\leq k_1},\left\{\zeta_{jn}\right\}_{1\leq n\leq k_2}\right) & ,\ \beta=2,\\
                        \left(\left\{\begin{array}{cc} z_{jn} & -z_{j+N,n}^* \\ z_{j+N,n} & z_{jn}^* \end{array}\right\}_{1\leq n\leq k_1},\displaystyle\left\{\begin{array}{cc} \zeta_{jn}^{(-)} & \zeta_{jn}^{(+)} \\ \zeta_{jn}^{(-)*} & \zeta_{jn}^{(+)*} \end{array}\right\}_{1\leq n\leq k_2}\right) & ,\ \beta=4,
                       \end{array}\right.
\end{equation}
respectively, where $\zeta_{jn}^{(\pm)}=\imath^{(1\pm1)/2}(\zeta_{jn}\pm\zeta_{j+N,n}^*)/\sqrt{2}$. Then, the supermatrix $\widehat{B}$ acquires the form
\begin{equation}\label{2.20}
 \widehat{B}=\frac{1}{\tilde{\gamma}}\sum\limits_{j=1}^N\Psi_j\Psi_j^\dagger .
\end{equation}
The integrand in Eq.~\eref{2.12}
\begin{equation}\label{2.21}
 F\left(\widehat{B}\right)=\Phi\left(\widehat{B}\right)\exp\left(-\imath\Str E\widehat{B}\right)
\end{equation}
comprises a symmetry breaking term,
\begin{equation}\label{2.22}
 \exists\ U\in\U^{(\beta)}(\gamma_1k_1/\gamma_2k_2)\ {\rm\ that\ \ }F\left(\widehat{B}\right)\neq F\left(U\widehat{B}U^\dagger\right) ,
\end{equation}
according to the supergroup
\begin{equation}\label{2.23}
 \U^{(\beta)}(\gamma_1k_1/\gamma_2k_2)=\left\{\begin{array}{ll}
                 \UOSp^{(+)}(2k_1/2k_2) & ,\ \beta=1\\
                 \U(k_1/k_2) & ,\ \beta=2\\
                 \UOSp^{(-)}(2k_1/2k_2) & ,\ \beta=4
                \end{array}\right. .
\end{equation}
We use the notation of Refs.~\cite{KohGuh05,KKG08} for the representations $\UOSp^{(\pm)}$ of the supergroup $\UOSp$. These representations are related to the classification of Riemannian symmetric superspaces by Zirnbauer \cite{Zir96}. The index ``$+$'' in Eq.~\eref{2.23} refers to real entries in the Boson--Boson block and to quaternionic entries in the Fermion--Fermion block and ``$-$'' indicates the other way around.

\section{Supersymmetric Wishart--matrices and some of their properties}\label{sec3}

We generalize the integrand \eref{2.21} to arbitrary sufficiently integrable superfunctions on rectangular $(\gamma_2c+\gamma_1d)\times(\gamma_2a+\gamma_1b)$ supermatrices $\widehat{V}$ on the complex Grassmann--algebra $\Lambda=\bigoplus\limits_{j=0}^{2(ad+bc)}\Lambda_j$. Such a supermatrix
\begin{equation}\label{3.1}
 \fl\widehat{V}=\left(\Psi_{11}^{({\rm C})},\ldots,\Psi_{a1}^{({\rm C})},\Psi_{12}^{({\rm C})},\ldots\Psi_{b2}^{({\rm C})}\right)=\left(\Psi_{11}^{({\rm R})*}\ldots,\Psi_{c1}^{({\rm R})*},\Psi_{12}^{({\rm R})*},\ldots\Psi_{d2}^{({\rm R})*}\right)^{T_{\rm S}}
\end{equation}
is defined by its columns
\begin{eqnarray}
 \Psi_{j1}^{({\rm C})\dagger} & = & \left\{\begin{array}{ll}
                        \left(\left\{x_{jn}\right\}_{1\leq n\leq c},\left\{\chi_{jn},\chi_{jn}^*\right\}_{1\leq n\leq d}\right) & ,\ \beta=1,\\
                        \left(\left\{z_{jn}\right\}_{1\leq n\leq c},\left\{\chi_{jn}\right\}_{1\leq n\leq d}\right) & ,\ \beta=2,\\
                        \left(\left\{\begin{array}{cc} z_{jn1} & -z_{jn2}^* \\ z_{jn2} & z_{jn1}^* \end{array}\right\}_{1\leq n\leq c},\left\{\begin{array}{c} \chi_{jn} \\ \chi_{jn}^* \end{array}\right\}_{1\leq n\leq d}\right) & ,\ \beta=4,
                       \end{array}\right.\label{3.2a}\\
 \Psi_{j2}^{({\rm C})\dagger} & = & \left\{\begin{array}{ll}
                        \left(\left\{\begin{array}{c} \zeta_{jn} \\ \zeta_{jn}^* \end{array}\right\}_{1\leq n\leq c},\left\{\begin{array}{cc} \tilde{z}_{jn1} & -\tilde{z}_{jn2}^* \\ \tilde{z}_{jn2} & \tilde{z}_{jn1}^* \end{array}\right\}_{1\leq n\leq d}\right) & ,\ \beta=1,\\
                        \left(\left\{\zeta_{jn}\right\}_{1\leq n\leq c},\left\{\tilde{z}_{jn}\right\}_{1\leq n\leq d}\right) & ,\ \beta=2,\\
                        \left(\left\{\zeta_{jn},\zeta^*_{jn}\right\}_{1\leq n\leq c},\left\{y_{jn}\right\}_{1\leq n\leq d}\right) & ,\ \beta=4,
                       \end{array}\right.\label{3.2b}
\end{eqnarray}
or by its rows
\begin{eqnarray}
 \Psi_{j1}^{({\rm R})\dagger} & = & \left\{\begin{array}{ll}
                        \left(\left\{x_{nj}\right\}_{1\leq n\leq a},\left\{ \zeta_{nj}^*, -\zeta_{nj} \right\}_{1\leq n\leq b}\right) & ,\ \beta=1,\\
                        \left(\left\{z_{nj}^*\right\}_{1\leq n\leq a},\left\{\zeta_{nj}^*\right\}_{1\leq n\leq b}\right) & ,\ \beta=2,\\
                        \left(\left\{\begin{array}{cc} z_{nj1}^* & z_{nj2}^* \\ -z_{nj2} & z_{nj1} \end{array}\right\}_{1\leq n\leq a},\left\{\begin{array}{c} \zeta_{nj}^* \\ -\zeta_{nj} \end{array}\right\}_{1\leq n\leq b}\right) & ,\ \beta=4,
                       \end{array}\right.\label{3.3a}\\
 \Psi_{j2}^{({\rm R})\dagger} & = & \left\{\begin{array}{ll}
                        \left(\left\{\begin{array}{c} -\chi_{nj}^* \\ \chi_{nj} \end{array}\right\}_{1\leq n\leq a},\left\{\begin{array}{cc} \tilde{z}_{nj1}^* & \tilde{z}_{nj2}^* \\ -\tilde{z}_{nj2} & \tilde{z}_{nj1} \end{array}\right\}_{1\leq n\leq b}\right) & ,\ \beta=1,\\
                        \left(\left\{-\chi_{nj}^*\right\}_{1\leq n\leq a},\left\{\tilde{z}_{nj}^*\right\}_{1\leq n\leq b}\right) & ,\ \beta=2,\\
                        \left(\left\{-\chi_{nj}^*,\chi_{nj}\right\}_{1\leq n\leq a},\left\{y_{nj}\right\}_{1\leq n\leq b}\right) & ,\ \beta=4
                       \end{array}\right.\label{3.3b}
\end{eqnarray}
which are real, complex and quaternionic supervectors. We use the complex Grassmann variables $\chi_{mn}$ and $\zeta_{mn}$ and the real numbers $x_{mn}$ and $y_{mn}$. Also, we introduce the complex numbers $z_{mn}$, $\tilde{z}_{mn}$, $z_{mnl}$ and $\tilde{z}_{mnl}$. The $(\gamma_2c+\gamma_1d)\times(\gamma_2c+\gamma_1d)$ supermatrix $\widehat{B}=\tilde{\gamma}^{-1}\widehat{V}\widehat{V}^\dagger$ can be written in the columns of $\widehat{V}$ as in Eq.~\eref{2.20}. As this supermatrix has a form similar to the ordinary Wishart--matrices, we refer to it as supersymmetric Wishart--matrix. The rectangular supermatrix above fulfills the property
\begin{equation}\label{3.4}
 \widehat{V}^*=\widehat{Y}_{cd}\widehat{V}\widehat{Y}_{ab}^T .
\end{equation}
The corresponding generating function \eref{2.1} is an integral over a rotation invariant superfunction $P$ on a superspace, which is sufficiently convergent and analytic in its real independent variables,
\begin{equation}\label{3.5}
  Z_{cd}^{ab}(E^-) =\int\limits_{\widetilde{\Sigma}_{\beta,ab}^{(-\psi)}} P(\sigma) \Sdet^{-1/\tilde{\gamma}}\left(\sigma\otimes\widehat{\Pi}_{2\psi}^{({\rm C})}- \eins_{\gamma_2a+\gamma_1b}\otimes E^-\right)d[\sigma],
\end{equation}
where
\begin{equation}\label{3.5b}
 \fl E^-=\diag\left(E_{11}\otimes \eins_{\gamma_2},\ldots,E_{c1}\otimes \eins_{\gamma_2},E_{12}\otimes \eins_{\gamma_1},\ldots,E_{d2}\otimes \eins_{\gamma_1}\right)-\imath\varepsilon \eins_{\gamma_2c+\gamma_1d} .
\end{equation}
Let $\widetilde{\Sigma}_{\beta,ab}^{0(\dagger)}$ be a subset of $\widetilde{\Sigma}_{\beta,ab}^{(\dagger)}$. The entries of elements in $\widetilde{\Sigma}_{\beta,ab}^{0(\dagger)}$ lie in $\Lambda_0$ and $\Lambda_1$. The set $\widetilde{\Sigma}_{\beta,ab}^{(-\psi)}=\widehat{\Pi}_{-\psi}^{({\rm R})}\widetilde{\Sigma}_{\beta,ab}^{0(\dagger)}\widehat{\Pi}_{-\psi}^{({\rm R})}$ is the Wick--rotated set of $\widetilde{\Sigma}_{\beta,ab}^{0(\dagger)}$ by the generalized Wick--rotation $\widehat{\Pi}_{-\psi}^{({\rm R})}=\diag( \eins_{\gamma_2a},e^{-\imath\psi/2} \eins_{\gamma_1b})$. As in Ref.~\cite{KGG08}, we introduce such a rotation for the convergence of the integral \eref{3.5}. The matrix $\widehat{\Pi}_{2\psi}^{({\rm C})}=\diag( \eins_{\gamma_2c},e^{\imath\psi} \eins_{\gamma_1d})$ is also a Wick--rotation.

In the rest of our work, we restrict the calculations to a class of superfunctions. These superfunctions has a Wick--rotation such that the integrals are convergent. We have not explicitly analysed the class of such functions. However, this class is very large and sufficient for physical interests. We consider the probability distribution
\begin{equation}\label{3.5c}
 P(\sigma)=f(\sigma)\exp(-\Str\sigma^{2m}),
\end{equation}
where $m\in\mathbb{N}$ and $f$ is a superfunction which does not increase so fast as $\exp(\Str\sigma^{2m})$ in the infinty, in particular
\begin{equation}\label{3.5d}
 \underset{\epsilon\to\infty}{\lim}P\left(\epsilon e^{\imath\alpha}\sigma\right)=0 \Leftrightarrow \underset{\epsilon\to\infty}{\lim}\exp\left(-\epsilon e^{\imath\alpha}\Str\sigma^{2m}\right)=0
\end{equation}
for every angle $\alpha\in[0,2\pi]$. Then, a Wick--rotation exists for $P$.

To guarantee the convergence of the integrals below, let $\widehat{V}_{\psi}=\widehat{\Pi}_{\psi}^{({\rm C})}\widehat{V}$, $\widehat{V}_{-\psi}^\dagger=\widehat{V}^\dagger\widehat{\Pi}_{\psi}^{({\rm C})}$ and $\widehat{B}_{\psi}=\widehat{\Pi}_{\psi}^{({\rm C})}\widehat{B}\widehat{\Pi}_{\psi}^{({\rm C})}$. Considering a function $f$ on the set of supersymmetric Wishart--matrices, we give a lemma and a corollary which are of equal importance for the superbosonization formula and the generalized Hubbard--Startonovich transformation. For $b=0$, the lemma presents the duality relation between the ordinary and superspace \eref{2.11} which is crucial for the calculation of \eref{2.1}. This lemma was proven in Ref.~\cite{LSZ07} by representation theory. Here, we only state it.
\begin{lemma}\label{c1}\ \\
 Let $f$ be a superfunction on rectangular supermatrices of the form \eref{3.1} and invariant under
 \begin{equation}\label{c1.1}
  f(\widehat{V}_{\psi},\widehat{V}_{-\psi}^\dagger)=f\left(\widehat{V}_{\psi}U^\dagger,U\widehat{V}_{-\psi}^\dagger\right)\ ,
 \end{equation}
 for all $\widehat{V}$ and $U\in\U^{(\beta)}(a/b)$. Then, there is a superfunction $F$ on the $\U^{(\beta)}(c/d)$--symmetric supermatrices with
 \begin{equation}\label{c1.2}
  F(\widehat{B}_{\psi})=f(\widehat{V}_{\psi},\widehat{V}_{-\psi}^\dagger)\ .
 \end{equation}
\end{lemma}

The $\U^{(\beta)}(c/d)$--symmetric supermatrices are elements of $\widetilde{\Sigma}_{\beta,ab}^{(\dagger)}$. The invariance condition \eref{c1.1} implies that $f$ only depends on the rows of $\widehat{V}_{\psi}$ by $\Psi_{nr}^{({\rm R})\dagger}\Psi_{ms}^{({\rm R})}$ for arbitrary $n,m,r$ and $s$. These scalar products are the entries of the supermatrix $\widehat{V}_{\psi}\widehat{V}_{-\psi}^\dagger$ which leads to the statement.

The corollary below is an application of integral theorems by Wegner \cite{Weg83} worked out in Refs.~\cite{Con88,ConGro89} and of the Theorems III.1, III.2 and III.3 in Ref.~\cite{KKG08}. It states that an integration over supersymmetric Wishart--matrices can be reduced to integrations over supersymmetric Wishart--matrices comprising a lower dimensional rectangular supermatrix. In particular for the generating function, it reflects the equivalence of the integral \eref{3.5} with an integration over smaller supermatrices \cite{KKG08}. We assume that $\tilde{a}=a-2(b-\tilde{b})/\beta\geq0$ with
\begin{equation}\label{3.6}
 \tilde{b}=\left\{\begin{array}{ll}
		1 & ,\ \beta=4\ {\rm and}\ b\in2\mathbb{N}_0+1\\
		0 & ,\ {\rm else}
	   \end{array}\right. .
\end{equation}
\begin{corollary}\label{c2}\ \\
 Let $F$ be the superfunction of Lemma \ref{c1}, real analytic in its real independent entries and a Schwartz--function. Then, we find
 \begin{equation}\label{c2.1}
  \displaystyle\int\limits_{\mathfrak{R}}F(\widehat{B}_{\psi})d[\widehat{V}]=C\int\limits_{\widetilde{\mathfrak{R}}}F(\widetilde{B}_{\psi})d[\widetilde{V}]
 \end{equation}
 where $\widetilde{B}=\tilde{\gamma}^{-1}\widetilde{V}\widetilde{V}$. The sets are $\mathfrak{R}=\mathbb{R}^{\beta ac+4bd/\beta}\times\Lambda_{2(ad+bc)}$ and $\widetilde{\mathfrak{R}}=\mathbb{R}^{\beta\tilde{a}c+4\tilde{b}d/\beta}\times\Lambda_{2(\tilde{a}d+\tilde{b}c)}$, the constant is
 \begin{equation}\label{c2.2}
  C=\left[-\frac{\gamma_1}{2}\right]^{(b-\tilde{b})c}\left[\frac{\gamma_2}{2}\right]^{(a-\tilde{a})d}
 \end{equation}
 and the measure
\begin{equation}\label{c2.3}
 d[\widehat{V}]=\underset{1\leq l\leq \beta}{\underset{1\leq n\leq c}{\underset{1\leq m\leq a}{\prod}}}dx_{mnl}\underset{1\leq l\leq 4/\beta}{\underset{1\leq n\leq d}{\underset{1\leq m\leq b}{\prod}}}dy_{mnl}\underset{1\leq n\leq c}{\underset{1\leq m\leq b}{\prod }}d\zeta_{mn}d\zeta_{mn}^*\underset{1\leq n\leq d}{\underset{1\leq m\leq a}{\prod }}d\chi_{mn}d\chi_{mn}^*\ .
\end{equation}
The $(\gamma_2c+\gamma_1d)\times(\gamma_2\tilde{a}+\gamma_1\tilde{b})$ supermatrix $\widetilde{V}$ and its measure $d[\widetilde{V}]$ is defined analogous to $\widehat{V}$ and $d[\widehat{V}]$, respectively. Here, $x_{mna}$ and $y_{mna}$ are the independent real components of the real, complex and quaternionic numbers of the supervectors $\Psi_{j1}^{({\rm R})}$ and $\Psi_{j2}^{({\rm R})}$, respectively.
\end{corollary}
\textbf{Proof:}\\
We integrate $F$ over all supervectors $\Psi_{j1}^{({\rm R})}$ and $\Psi_{j2}^{({\rm R})}$ except $\Psi_{11}^{({\rm R})}$. Then, 
 \begin{equation}\label{c2.4}
  \displaystyle\int\limits_{\mathfrak{R}^\prime}F(V_{\psi}V_{-\psi}^\dagger)d[\widehat{V}_{\neq11}]
 \end{equation}
only depends on $\Psi_{11}^{({\rm R})\dagger}\Psi_{11}^{({\rm R})}$. The integration set is $\mathfrak{R}^\prime=\mathbb{R}^{\beta a(c-1)+4bd/\beta}\times\Lambda_{2(ad+b(c-1))}$ and the measure $d[\widehat{V}_{\neq11}]$ is $d[\widehat{V}]$ without the measure for the supervector $\Psi_{11}^{({\rm R})}$. With help of the Theorems in Ref.~\cite{Weg83,Con88,ConGro89,KKG08}, the integration over $\Psi_{11}^{({\rm R})}$ is up to a constant equivalent to an integration over a supervector $\widetilde{\Psi}_{11}^{({\rm R})}$. This supervector is equal to $\Psi_{11}^{({\rm R})}$ in the first $\tilde{a}$--th entries and else zero. We repeat this procedure for all other supervectors reminding that we only need the invariance under the supergroup action $\U^{(\beta)}\left(b-\tilde{b}/b-\tilde{b}\right)$ on $f$ as in Eq.~\eref{c1.1} embedded in $\U^{(\beta)}(a/b)$. This invariance is preserved in each step due to the zero entries in the new supervectors. \hfill$\square$

This corollary allows us to restrict our calculation on supermatrices with $b=1$ only to $\beta=4$ and $b=0$ for all $\beta$. Only the latter case is of physical interest. Thus, we give the computation for $b=0$ in the following sections and consider the case $b=1$ in Sec. \ref{sec7}. For $b=0$ we omit the Wick--rotation for $\widehat{B}$ as it is done in Refs.~\cite{Guh06,KGG08} due to the convergence of the integral \eref{3.5}.

\section{The superbosonization formula}\label{sec4}

We need for the following theorem the definition of the sets
\begin{eqnarray}
 \fl\Sigma_{1,pq} = \left\{\left.\sigma=\left[\begin{array}{ccc} \sigma_1 & \eta & \eta^* \\ -\eta^\dagger & \sigma_{21} & \sigma_{22}^{(1)} \\ \eta^T  & \sigma_{22}^{(2)} & \sigma_{21}^T \end{array}\right]\in{\rm Mat}(p/2q)\right|\sigma_1^\dagger=\sigma_1^*=\sigma_1\ {\rm with\ positive}\right.\nonumber\\
 \left.{\rm definite\ body},\ \sigma_{22}^{(1)T}=-\sigma_{22}^{(1)},\ \sigma_{22}^{(2)T}=-\sigma_{22}^{(2)}\right\},\label{4.1}\\
 \fl\Sigma_{2,pq} = \left\{\left.\sigma=\left[\begin{array}{cc} \sigma_1 & \eta \\ -\eta^\dagger & \sigma_2 \end{array}\right]\in{\rm Mat}(p/q)\right|\sigma_1^\dagger=\sigma_1\ {\rm with\ positive\ definite\ body}\right\},\label{4.2}\\
 \fl\Sigma_{4,pq} = \left\{\sigma=\left[\begin{array}{ccc} \sigma_{11} & \sigma_{12} & \eta \\ -\sigma_{12}^* & \sigma_{11}^* & \eta^* \\ -\eta^\dagger  & \eta^T & \sigma_2 \end{array}\right]\in{\rm Mat}(2p/q)\right|\sigma_1^\dagger=\sigma_1=\left[\begin{array}{cc} \sigma_{11} & \sigma_{12} \\ -\sigma_{12}^* & \sigma_{11}^* \end{array}\right]\nonumber\\
 {\rm with\ positive\ definite\ body},\ \sigma_2=\sigma_2^T\Biggl\}.\label{4.3}
\end{eqnarray}
Also, we will use the sets
\begin{equation}\label{4.4}
 \Sigma_{\beta,pq}^{(\dagger)} = \left\{\left.\sigma\in\Sigma_{\beta,pq}\right|\sigma_2^\dagger=\sigma_2\right\}=\widetilde{\Sigma}_{\beta,pq}^{(\dagger)}\cap\Sigma_{\beta,pq}
\end{equation}
and
\begin{equation}\label{4.5}
 \Sigma_{\beta,pq}^{({\rm c})} = \left\{\left.\sigma\in\Sigma_{\beta,pq}\right|\sigma_2\in\CU^{(4/\beta)}\left(q\right)\right\}
\end{equation}
where $\CU^{(\beta)}\left(q\right)$ is the set of the circular orthogonal (COE, $\beta=1$), unitary (CUE, $\beta=2$) or unitary-symplectic (CSE, $\beta=4$) ensembles,
\begin{equation}\label{4.6}
 \fl\CU^{(\beta)}\left(q\right)=\left\{A\in{\rm Gl}(\gamma_2q,\mathbb{C})\left| \begin{array}{ll} A=A^T\in\U^{(2)}(q) & ,\ \beta=1 \\ A\in\U^{(2)}(q) & ,\ \beta=2 \\ A=(Y_s\otimes\eins_q)A^T(Y_s^T\otimes\eins_q)\in\U^{(2)}(2q) & ,\ \beta=4 \end{array}\right.\right\}
\end{equation}
The index ``$\dagger$'' in Eq.~\eref{4.4} refers to the self-adjointness of the supermatrices and the index ``${\rm c}$'' indicates the relation to the circular ensembles. We notice that the set classes presented above differ in the Fermion--Fermion block. In Sec. \ref{sec6}, we show that this is the crucial difference between both methods. Due to the nilpotence of $B$'s Fermion--Fermion block, we can change the set in this block for the Fourier--transformation. The sets of matrices in the sets above with entries in $\Lambda_0$ and $\Lambda_1$ are denoted by $\Sigma_{\beta,pq}^{0}$, $\Sigma_{\beta,pq}^{0(\dagger)}$ and $\Sigma_{\beta,pq}^{0({\rm c})}$, respectively.

The proof of the superbosonization formula \cite{Som07,LSZ07} given below is based on the proofs of the superbosonization formula for arbitrary superfunctions on real supersymmetric Wishart--matrices in Ref.~\cite{Som07} and for Gaussian functions on real, complex and quaternionic Wishart--matrices in Ref.~\cite{Som08}. This theorem extends the superbosonization formula of Ref.~\cite{LSZ07} to averages of square roots of determinants over unitary-symplectically invariant ensembles, i.e. $\beta=4$, $b=c=0$ and $d$ odd in Eq.~\eref{3.5}. The proof of this theorem is given in  \ref{app1}.
\begin{theorem}[Superbosonization formula]\label{t1}\ \\
 Let $F$ be a conveniently integrable and analytic superfunction on the set of $\left(\gamma_2c+\gamma_1d\right)\times\left(\gamma_2c+\gamma_1d\right)$ supermatrices and
\begin{equation}\label{t1.3}
 \kappa=\frac{a-c+1}{\gamma_1}+\frac{d-1}{\gamma_2} .
\end{equation}
With
\begin{equation}\label{t1.0}
 a\geq c\ ,
\end{equation}
we find
 \begin{equation}\label{t1.1}
  \fl\int\limits_{\mathfrak{R}}F(\widehat{B})\exp\left(-\varepsilon\Str \widehat{B}\right)d[\widehat{V}]=C_{acd}^{(\beta)}\int\limits_{\Sigma_{\beta,cd}^{0({\rm c})}}F(\rho)\exp\left(-\varepsilon\Str\rho\right)\Sdet\rho^{\kappa}d[\rho] ,
 \end{equation}
 where the constant is
\begin{eqnarray}
 \fl C_{acd}^{(\beta)}
 =  \left(-2\pi\gamma_1\right)^{-ad}\left(-\frac{2\pi}{\gamma_2}\right)^{cd}2^{-c}\tilde{\gamma}^{\beta ac/2}\frac{{\rm Vol}\left(\U^{(\beta)}(a)\right)}{{\rm Vol}\left(\U^{(\beta)}(a-c)\right)}\times\nonumber\\
 \times\prod\limits_{n=1}^d\frac{\Gamma\left(\gamma_1\kappa+2(n-d)/\beta\right)}{\imath^{4(n-1)/\beta}\pi^{2(n-1)/\beta}} .\label{t1.2}
\end{eqnarray}
We define the measure $d[\widehat{V}]$ as in Corollary \ref{c2} and the measure on the right hand side is $d[\rho]=d[\rho_1]d[\rho_2]d[\eta]$ where
\begin{eqnarray}
 \fl d[\rho_1] & = & \prod\limits_{n=1}^{c}d\rho_{nn1}\times\left\{\begin{array}{ll}
                  \prod\limits_{1\leq n< m\leq c}d\rho_{nm1} & ,\ \beta=1,\\
                  \prod\limits_{1\leq n< m\leq c}d\RE\rho_{nm1}d\IM\rho_{nm1} & ,\ \beta=2,\\
                  \prod\limits_{1\leq n< m\leq c}d\RE\rho_{nm11}d\IM\rho_{nm11}d\RE\rho_{nm12}d\IM\rho_{nm12} & ,\ \beta=4,
                 \end{array}\right.\label{t1.4}\\
 \fl d[\rho_2] & = & {\rm FU}_{d}^{(4/\beta)}|\Delta_{d}(e^{\imath\varphi_j})|^{4/\beta}\prod\limits_{n=1}^{d}\frac{de^{\imath\varphi_n}}{2\pi\imath}d\mu(U)\label{t1.5} ,\\
 \fl d[\eta]    & = & \prod\limits_{n=1}^{c}\prod\limits_{m=1}^{d}(d\eta_{nm}d\eta_{nm}^*)\label{t1.6}.
\end{eqnarray}
Here, $\rho_2=U\diag\left(e^{\imath\varphi_1},\ldots,e^{\imath\varphi_{d}}\right)U^\dagger$, $U\in\U^{(4/\beta)}\left(d\right)$ and $d\mu(U)$ is the normalized Haar-measure of $\U^{(4/\beta)}\left(d\right)$. We introduce the volumes of the rotation groups
\begin{equation}\label{t1.7}
 {\rm Vol}\left(\U^{(\beta)}(n)\right)=\prod\limits_{j=1}^n\frac{2\pi^{\beta j/2}}{\Gamma\left(\beta j/2\right)}
\end{equation}
and the ratio of volumes of the group flag manifold and the permutation group
\begin{equation}\label{t1.8}
 {\rm FU}_{d}^{(4/\beta)}=\frac{1}{d!}\prod\limits_{j=1}^d\frac{\pi^{2(j-1)/\beta}\Gamma(2/\beta)}{\Gamma(2j/\beta)}\ .
\end{equation}
The absolute value of the Vandermonde determinant $\Delta_{d}(e^{\imath\varphi_j})=\prod\limits_{1\leq n<m\leq d}\left(e^{\imath\varphi_n}-e^{\imath\varphi_m}\right)$ refers to a change of sign in every single difference $\left(e^{\imath\varphi_n}-e^{\imath\varphi_m}\right)$ with ``$+$'' if $\varphi_m<\varphi_n$ and with ``$-$'' otherwise. Thus, it is not an absolute value in the complex plane.
\end{theorem}

The exponential term can also be shifted in the superfunction $F$. We need this additional term to regularize an intermediate step in the proof.

The inequality \eref{t1.0} is crucial. For example, let $\beta=2$ and $F(\rho)=1$. Then, the left hand side of Eq.~\eref{t1.1} is not equal to zero. On the right hand side of Eq.~\eref{t1.1}, the dependence on the Grassmann variables  only stems from the superdeterminant and we find
\begin{equation}\label{4.7}
  \int\limits_{\Lambda_{2cd}}\Sdet\rho^\kappa d[\eta]=\int\limits_{\Lambda_{2cd}}\frac{\det\left(\rho_1+\eta\rho_2^{-1}\eta^\dagger\right)^\kappa}{\det\rho_2^\kappa} d[\eta]=0
\end{equation}
for $\kappa<d$. The superdeterminant $\Sdet\rho$ is a polynomial of order $2c$ in the Grassmann variables $\{\eta_{nm},\eta_{nm}^*\}$ and the integral over the remaining variables is finite for $\kappa\geq0$. Hence, it is easy to see that the right hand side of Eq.~\eref{t1.1} is zero for $\kappa<d$. This inequality is equivalent to $a<c$.

This problem was also discussed in Ref.~\cite{BEKYZ07}. These authors gave a solution for the case that \eref{t1.0} is violated. This solution differs from our approach in Sec. \ref{sec7}.

\section{The generalized Hubbard--Stratonovich transformation}\label{sec5}

The following theorem is proven in a way similar to Refs.~\cite{Guh06,KGG08}. The proof is given in \ref{app2}. We need the Wick--rotated set $\Sigma_{\beta,cd}^{(\psi)}=\widehat{\Pi}_\psi^{({\rm C})}\Sigma_{\beta,cd}^{0(\dagger)}\widehat{\Pi}_\psi^{({\rm C})}$, particularly $\Sigma_{\beta,cd}^{(0)}=\Sigma_{\beta,cd}^{0(\dagger)}$. The original extension of the Hubbard--Stratonovich transformation \cite{Guh06,KGG08} was only given for $\gamma_2c=\gamma_1d=\tilde{\gamma}k$. Here, we generalize it to arbitrary $c$ and $d$.
\begin{theorem}[Generalized Hubbard--Stratonovich transformation]\ \label{t2}\\
 Let $F$ and $\kappa$ be the same as in Theorem \ref{t1}. If the inequality \eref{t1.0} holds, we have
 \begin{eqnarray}
   \fl\int\limits_{\mathfrak{R}}F(\widehat{B})\exp\left(-\varepsilon\Str \widehat{B}\right)d[\Psi]=\nonumber\\
   \fl= \widetilde{C}_{acd}^{(\beta)}\int\limits_{\Sigma_{\beta,cd}^{(\psi)}}F\left(\hat{\rho}\right)\exp\left(-\varepsilon\Str\hat{\rho}\right)\det\rho_1^\kappa \left(e^{-\imath\psi d}D_{dr_2}^{(4/\beta)}\right)^{a-c}\frac{\delta(r_2)}{|\Delta_d(e^{\imath\psi}r_2)|^{4/\beta}}e^{-\imath\psi cd}d[\rho]=\nonumber\\
    \fl= \widetilde{C}_{acd}^{(\beta)}\int\limits_{\Sigma_{\beta,cd}^{(0)}}\det\rho_1^\kappa\frac{\delta(r_2)}{|\Delta_d(r_2)|^{4/\beta}} \left((-1)^dD_{dr_2}^{(4/\beta)}\right)^{a-c}\left.F\left(\hat{\rho}\right)\exp\left(-\varepsilon\Str\hat{\rho}\right)\right|_{\psi=0}d[\rho]\label{t2.1}
 \end{eqnarray}
 with
 \begin{equation}\label{t2.0}
  \hat{\rho}=\left[\begin{array}{c|c} \rho_1 & e^{\imath\psi/2}\rho_\eta \\ \hline -e^{\imath\psi/2}\rho_\eta^\dagger & e^{\imath\psi}\left(\rho_2-\rho_\eta^\dagger\rho_1^{-1}\rho_\eta\right) \end{array}\right].
 \end{equation}
 The variables $r_2$ are the eigenvalues of the supermatrix $\rho_2$. The measure $d[\rho]=d[\rho_1]d[\rho_2]d[\eta]$ is defined by Eqs. \eref{t1.4} and \eref{t1.6}. For the measure $d[\rho_2]$ we take the definition \eref{t1.4} for $4/\beta$. The differential operator in Eq.~\eref{t2.1} is an analog of the Sekiguchi--differential operator \cite{OkoOls97} and has the form \cite{KGG08}
 \begin{equation}\label{t2.2}
  D_{dr_2}^{(4/\beta)}=\frac{1}{\Delta_d(r_2)}\det\left[r_{n2}^{d-m}\left(\frac{\partial}{\partial r_{n2}}+(d-m)\frac{2}{\beta}\frac{1}{r_{n2}}\right)\right]_{1\leq n,m\leq d} .
 \end{equation}
 The constant is
 \begin{equation}\label{t2.3}
  \widetilde{C}_{acd}^{(\beta)}=2^{-c}\left(2\pi\gamma_1\right)^{-ad}\left(\frac{2\pi}{\gamma_2}\right)^{cd}\tilde{\gamma}^{\beta ac/2}\frac{{\rm Vol}\left(\U^{(\beta)}(a)\right)}{{\rm Vol}\left(\U^{(\beta)}(a-c)\right){\rm FU}_d^{(4/\beta)}} .
 \end{equation}
\end{theorem}

Since the diagonalization of $\rho_2$ yields an $|\Delta_d(r_2)|^{4/\beta}$ in the measure, the ratio of the Dirac--distribution with the Vandermonde--determinant is for Schwartz--functions on $\Herm(4/\beta,d)$ well--defined. Also, the action of $D_{dr_2}^{(4/\beta)}$ on such a Schwartz--function integrated over the corresponding rotation group is finite at zero.

The distribution in the Fermion--Fermion block in Eq.~\eref{t2.1} takes for $\beta\in\{1,2\}$ the simpler form \cite{Guh06,KGG08}
\begin{eqnarray}
 \fl\left(D_{dr_2}^{(4/\beta)}\right)^{a-c}\frac{\delta(r_2)}{|\Delta_d(r_2)|^{4/\beta}} =\nonumber\\
 ={\rm FU}_d^{(4/\beta)}\prod\limits_{n=1}^{d}\frac{\Gamma\left(a-c+1+2(n-1)/\beta \right)}{(-\pi)^{2(n-1)/\beta}\Gamma\left(\gamma_1\kappa\right)}\prod\limits_{n=1}^{d}\frac{\partial^{\gamma_1\kappa-1}}{\partial r_{n2}^{\gamma_1\kappa-1}}\delta(r_{2n})\label{5.1}\ .
\end{eqnarray}
This expression written as a contour integral is the superbosonization formula \cite{BasAke07}. For $\beta=4$, we do not find such a simplification due to the term $|\Delta(r_2)|$ as the Jacobian in the eigenvalue--angle coordinates.

\section{Equivalence of and connections between the two approaches}\label{sec6}

Above, we have argued that both expressions in Theorems \ref{t1} and \ref{t2} are equivalent for $\beta\in\{1,2\}$. Now we address all $\beta\in\{1,2,4\}$. The Theorem below is proven in  \ref{app3}. The proof treats all three cases in a unifying way. Properties of the ordinary matrix Bessel--functions are used.
\begin{theorem}[Equivalence of Theorems \ref{t1} and \ref{t2}]\ \label{t3}\\
 The superbosonization formula, \ref{t1}, and the generalized Hubbard--Stratonovich transformation, \ref{t2}, are equivalent for superfunctions which are Schwartz--functions and analytic in the fermionic eigenvalues.
\end{theorem}

\begin{figure}[ht]
 \centering
 \includegraphics[width=0.7\textwidth]{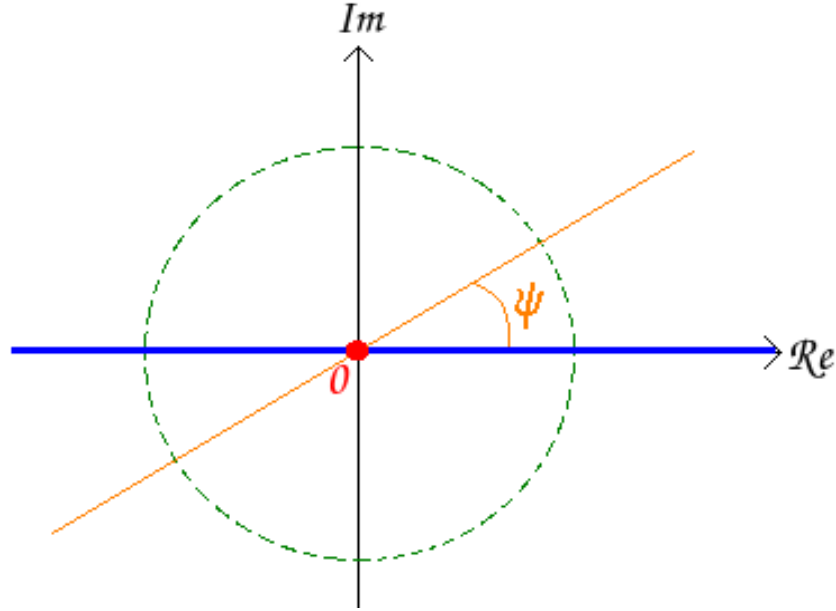}
 \caption{In the superbosonization formula, the integration of the fermionic eigenvalues is along the unit circle in the complex plane (dotted circle). The eigenvalue integrals in the generalized Hubbard--Stratonovich transformation are integrations over the real axis (bold line) or on the Wick--rotated real axis (thin line at angle $\psi$) if the differential operator acts on the superfunction or on the Dirac--distribution at zero (bold dot, $0$), respectively.}\label{f1}
\end{figure}

The compact integral in the Fermion--Fermion block of the superbosonization formula can be considered as a contour integral. In the proof of Theorem \ref{t3}, we find the integral identity
\begin{eqnarray}
  & & \int\limits_{[0,2\pi]^{d}}\widetilde{F}\left(e^{\imath\varphi_j}\right)\left|\Delta_d\left(e^{\imath\varphi_j}\right)\right|^{4/\beta}\prod\limits_{n=1}^d\frac{e^{\imath(1-\gamma_1\kappa)\varphi_n}d\varphi_n}{2\pi}=\nonumber\\
  & = & \prod\limits_{n=1}^d\frac{\imath^{4(n-1)/\beta}\Gamma(1+2n/\beta)}{\Gamma(2/\beta+1)\Gamma(\gamma_1\kappa-2(n-1)/\beta)}\left.\left(D_{dr_2}^{(4/\beta)}\right)^{a-c}\widetilde{F}(r_2)\right|_{r_2=0}\label{6.1}
\end{eqnarray}
for an analytic function $\widetilde{F}$ on $\mathbb{C}^d$ with permutation invariance. Hence, we can relate both constants \eref{t1.2} and \eref{t2.3},
\begin{equation}\label{6.2}
 \frac{\widetilde{C}_{acd}^{(\beta)}}{C_{acd}^{(\beta)}}=(-1)^{d(a-c)}\prod\limits_{n=1}^d\frac{\imath^{4(n-1)/\beta}\Gamma(1+2n/\beta)}{\Gamma(2/\beta+1)\Gamma(\gamma_1\kappa-2(n-1)/\beta)}.
\end{equation}
The integral identity \eref{6.1} is a reminiscent of the residue theorem. It is the analog of the connection between the contour integral and the differential operator in the cases $\beta\in\{1,2\}$, see Fig. \ref{f1}. Thus, the differential Operator with the Dirac--distribution in the generalized Hubbard--Stratonovich transformation restricts the non-compact integral in the Fermion--Fermion block to the point zero and its neighborhood. Therefore it is equivalent to a compact Fermion--Fermion block integral as appearing in the superbosonization formula.

\section{The general case for arbitrary positive integers $a$, $b$, $c$, $d$ and arbitrary Dyson--index $\beta\in\{1,2,4\}$}\label{sec7}

We consider an application of our results. The inequality \eref{t1.0} reads
\begin{equation}\label{7.1}
  N\geq\gamma_1k
\end{equation}
for the calculation of the $k$--point correlation function \eref{2.3} with help of the matrix Green function. For $\beta=1$ , a $N\times N$ real symmetric matrix has in the absence of degeneracies $N$ different eigenvalues. However, we can only calculate $k$--point correlation functions with $k<N/2$. For $N\to\infty$, this restriction does not matter. But for exact finite $N$ calculations, we have to modify the line of reasoning.

We construct the symmetry operator
\begin{equation}\label{7.2}
 \mathfrak{S}\left(\sigma\right)=\mathfrak{S}\left(\left[\begin{array}{cc} \sigma_{11} & \sigma_{12} \\ \sigma_{21} & \sigma_{22} \end{array}\right]\right)=\left[\begin{array}{cc} -\sigma_{22} & -\sigma_{21} \\ \sigma_{12} & \sigma_{11} \end{array}\right]
\end{equation}
from $(m_1+m_2)\times(n_1+n_2)$ supermatrix to $(m_2+m_1)\times(n_2+n_1)$ supermatrix. This operator has the properties
\begin{eqnarray}
 \mathfrak{S}(\sigma^\dagger) & = & \mathfrak{S}(\sigma)^\dagger\quad, \label{7.3}\\
 \mathfrak{S}(\sigma^*) & = & \mathfrak{S}(\sigma)^*, \label{7.4}\\
 \mathfrak{S}^2(\sigma) & = & -\sigma, \label{7.5}\\
 \fl\mathfrak{S}\left(\left[\begin{array}{cc} \sigma_{11} & \sigma_{12} \\ \sigma_{21} & \sigma_{22} \end{array}\right]\left[\begin{array}{cc} \rho_{11} & \rho_{12} \\ 0 & 0 \end{array}\right]\right) & = & \mathfrak{S}\left(\left[\begin{array}{cc} \sigma_{11} & \sigma_{12} \\ \sigma_{21} & \sigma_{22} \end{array}\right]\right)\mathfrak{S}\left(\left[\begin{array}{cc} \rho_{11} & \rho_{12} \\ 0 & 0 \end{array}\right]\right). \label{7.6}
\end{eqnarray}

Let $a,\ b,\ c$, $d$ be arbitrary positive integers and $\beta\in\{1,2,4\}$. Then, the equation \eref{7.6} reads for a matrix product of a $(\gamma_2c+\gamma_1d)\times(0+\gamma_1b)$ supermatrix with a $(0+\gamma_1b)\times(\gamma_2c+\gamma_1d)$ supermatrix
\begin{equation}\label{7.7}
 \fl\left[\begin{array}{c} \zeta^\dagger \\ \tilde{z}^\dagger \end{array}\right]\left[\begin{array}{cc} \zeta & \tilde{z} \end{array}\right]=\mathfrak{S}\left(\left[\begin{array}{c} \tilde{z}^\dagger \\ -\zeta^\dagger \end{array}\right]\right)\mathfrak{S}\left(\left[\begin{array}{cc} \tilde{z} & \zeta \end{array}\right]\right)=\mathfrak{S}\left(\left[\begin{array}{c} \tilde{z}^\dagger \\ -\zeta^\dagger \end{array}\right]\left[\begin{array}{cc} \tilde{z} & \zeta \end{array}\right]\right).
\end{equation}
With help of the operator $\mathfrak{S}$, we split the supersymmetric Wishart--matrix $\widehat{B}$ into two parts,
\begin{equation}\label{7.8}
 \widehat{B}=\widehat{B}_1+\mathfrak{S}(\widehat{B}_2)
\end{equation}
such that
\begin{equation}\label{7.9}
 \fl\widehat{B}_1=\tilde{\gamma}^{-1}\sum\limits_{j=1}^{a}\Psi_{j1}^{({\rm C})}\Psi_{j1}^{({\rm C})\dagger}\qquad{\rm and}\qquad\widehat{B}_2=\tilde{\gamma}^{-1}\sum\limits_{j=1}^{b}\mathfrak{S}\left(\Psi_{j2}^{({\rm C})}\right)\mathfrak{S}\left(\Psi_{j2}^{({\rm C})}\right)^\dagger .
\end{equation}
The supervectors $\mathfrak{S}\left(\Psi_{j2}^{({\rm C})}\right)$ are of the same form as $\Psi_{j1}^{({\rm C})}$. Let $\sigma$ be a quadratic supermatrix, i.e. $m_1=n_1$ and $m_2=n_2$. Then, we find the additional property
\begin{equation}\label{7.10}
 \Sdet\mathfrak{S}(\sigma)=(-1)^{m_2}\Sdet^{-1}\sigma .
\end{equation}

Let $\widehat{\Sigma}_{\beta,pq}^{(0)}=\mathfrak{S}\left(\Sigma_{\beta,pq}^{(0)}\right)$, $\widehat{\Sigma}_{\beta,pq}^{0({\rm c})}=\mathfrak{S}\left(\Sigma_{\beta,pq}^{0({\rm c})}\right)$ and the Wick--rotated set $\widehat{\Sigma}_{\beta,pq}^{(\psi)}=\widehat{\Pi}_{\psi}^{({\rm C})}\widehat{\Sigma}_{\beta,pq}^{(0)}\widehat{\Pi}_{\psi}^{({\rm C})}$. Then, we construct the analog of the superbosonization formula and the generalized Hubbard--Stratonovich transformation.
\begin{theorem}\label{t4}\ \\
 Let $F$ be the superfunction as in Theorem \ref{t1} and 
 \begin{equation}\label{t4.1}
  \kappa=\frac{a-c+1}{\gamma_1}-\frac{b-d+1}{\gamma_2} .
 \end{equation}
 Also, let $e\in\mathbb{N}_0$ and
 \begin{equation}\label{t4.2}
 \tilde{a}=a+\gamma_1e\qquad{\rm and}\qquad\tilde{b}=b+\gamma_2e
 \end{equation}
 with
 \begin{equation}\label{t4.3}
 \tilde{a}\geq c\qquad\tilde{b}\geq d .
 \end{equation}
 We choose the Wick--rotation  $e^{\imath\psi}$ such that all integrals are convergent. Then, we have
 \begin{eqnarray}
   \fl\int\limits_{\mathfrak{R}}F(\widehat{B}_{\psi})\exp\left(-\varepsilon\Str \widehat{B}_{\psi}\right)d[\widehat{V}]=\nonumber\\
   \fl= \left(-\frac{2}{\gamma_1}\right)^{\gamma_2ec}\left(\frac{2}{\gamma_2}\right)^{\gamma_1ed} \int\limits_{\widetilde{\mathfrak{R}}}F(\widetilde{B}_{\psi})\exp\left(-\varepsilon\Str \widetilde{B}_{\psi}\right)d[\widetilde{V}]=\nonumber\\
  \fl= C_{{\rm SF}}\int\limits_{\Sigma_{\beta,cd}^{0({\rm c})}}\int\limits_{\widehat{\Sigma}_{4/\beta,dc}^{0({\rm c})}}d[\rho^{(2)}]d[\rho^{(1)}]F(\rho^{(1)}+e^{\imath\psi}\rho^{(2)})\exp\left[-\varepsilon\Str(\rho^{(1)}+e^{\imath\psi}\rho^{(2)})\right]\times\nonumber\\
  \fl\times \Sdet^{\kappa+\tilde{b}/\gamma_2}\rho^{(1)}\Sdet^{\kappa-\tilde{a}/\gamma_1}\rho^{(2)}=\label{t4.5a}\\
  \fl =C_{{\rm HS}}\int\limits_{\Sigma_{\beta,cd}^{(0)}}\int\limits_{\widehat{\Sigma}_{4/\beta,cd}^{(0)}}d[\rho^{(2)}]d[\rho^{(1)}]\frac{\delta\left(r_2^{(1)}\right)}{\left|\Delta_d\left(r_2^{(1)}\right)\right|^{4/\beta}}\frac{\delta\left(r_1^{(2)}\right)}{\left|\Delta_c\left(r_1^{(2)}\right)\right|^{\beta}}{\det}^{\kappa+b/\gamma_2}\rho_1^{(1)}{\det}^{a/\gamma_1-\kappa}\rho_2^{(2)}\times\nonumber\\
  \fl\times \left(D_{dr_2^{(1)}}^{(4/\beta)}\right)^{\tilde{a}-c}\left(D_{cr_1^{(2)}}^{(\beta)}\right)^{\tilde{b}-d}F(\hat{\rho}^{(1)}+e^{\imath\psi}\hat{\rho}^{(2)})\exp\left[-\varepsilon\Str(\hat{\rho}^{(1)}+e^{\imath\psi}\hat{\rho}^{(2)})\right], \label{t4.5b}
 \end{eqnarray}
 where the constants are
 \begin{eqnarray}
  C_{{\rm SF}} & = & (-1)^{c(b-d)}e^{\imath\psi(\tilde{a}d-\tilde{b}c)}\left(\frac{2}{\gamma_1}\right)^{\gamma_2ec}\left(\frac{2}{\gamma_2}\right)^{\gamma_1ed}C_{\tilde{a}cd}^{(\beta)}C_{\tilde{b}dc}^{(4/\beta)} ,\\
  C_{{\rm HS}} & = & (-1)^{d(a-c)}e^{\imath\psi(\tilde{a}d-\tilde{b}c)}\left(-\frac{2}{\gamma_1}\right)^{\gamma_2ec}\left(-\frac{2}{\gamma_2}\right)^{\gamma_1ed}\widetilde{C}_{\tilde{a}cd}^{(\beta)}\widetilde{C}_{\tilde{b}dc}^{(4/\beta)} .
 \end{eqnarray}
 Here, we define the supermatrix
 \begin{equation}\label{t4.6}
  \fl\hat{\rho}^{(1)}+e^{\imath\psi}\hat{\rho}^{(2)}=\left[\begin{array}{c|c} \rho_1^{(1)}+e^{\imath\psi}\left(\rho_1^{(2)}-\rho_{\tilde{\eta}}^{(2)}\rho_2^{(2)-1}\rho_{\tilde{\eta}}^{(2)\dagger}\right) & \rho_\eta^{(1)}+e^{\imath\psi}\rho_{\tilde{\eta}}^{(2)} \\ \hline -\rho_\eta^{(1)\dagger}-e^{\imath\psi}\rho_{\tilde{\eta}}^{(2)\dagger} & \rho_2^{(1)}-\rho_\eta^{(1)\dagger}\rho_1^{(1)-1}\rho_\eta^{(1)}+e^{\imath\psi}\rho_2^{(2)} \end{array}\right]
 \end{equation}
 The set $\widetilde{\mathfrak{R}}$ is given as in corollary \ref{c2}. The measures $d[\rho^{(1)}]=d[\rho_1^{(1)}]d[\rho_2^{(1)}]d[\eta]$ and $d[\rho^{(2)}]=d[\rho_1^{(2)}]d[\rho_2^{(2)}]d[\tilde{\eta}]$ are given by Theorem \ref{t1}. The measures \eref{t1.4} for $\beta$ and $4/\beta$ assign $d[\rho_1^{(1)}]$ and $d[\rho_2^{(2)}]$ in Eqs. \eref{t4.5a} and \eref{t4.5b}, respectively. In Eq.~\eref{t4.5a}, $d[\rho_2^{(1)}]$ and $d[\rho_1^{(2)}]$ are defined by the measure \eref{t1.5} for the cases $\beta$ and $4/\beta$, respectively, and, in Eq.~\eref{t4.5b}, they are defined by the measure \eref{t1.4} for the cases $4/\beta$ and $\beta$, respectively. The measures $d[\eta]$ and $d[\tilde{\eta}]$ are the product of all complex Grassmann pairs as in Eq.~\eref{t1.6}.
\end{theorem}
Since this Theorem is a consequence of corollary \ref{c2} and Theorems \ref{t1} and \ref{t2}, the proof is quite simple.\\
\textbf{Proof:}\\
Let $e\in\mathbb{N}_0$ as in Eq.~\eref{t4.2}. Then, we use corollary \ref{c2} to extend the integral over $\widehat{V}$ to an integral over $\widetilde{V}$. We split the supersymmetric Wishart--matrix $\widehat{B}$ as in Eq.~\eref{7.8}. Both Wishart--matrices $\widehat{B}_{1}$ and $\widehat{B}_{2}$ fulfill the requirement \eref{t1.0} according to their dimension. Thus, we singly apply both Theorems \ref{t1} and \ref{t2} to $\widehat{B}_{1}$ and $\widehat{B}_{2}$. \hfill$\square$

Our approach of a violation of inequality \eref{t1.0} is quite different from the solution given in Ref.~\cite{BEKYZ07}. These authors introduce a matrix which projects the Boson--Boson block and the bosonic side of the off-diagonal blocks onto a space of the smaller dimension $a$. Then, they integrate over all of such orthogonal projectors. This integral becomes more difficult due to an additional measure on a curved, compact space. We use a second symmetric supermatrix. Hence, we have up to the dimensions of the supermatrices a symmetry between both supermatrices produced by $\mathfrak{S}$. There is no additional complication for the integration, since the measures of both supermatrices are of the same kind. Moreover, our approach extends the results to the case of $\beta=4$ and odd $b$ which is not considered in Ref.~\cite{BEKYZ07}.

\section{Remarks and conclusions}\label{sec8}

We proved the equivalence of the generalized Hubbard--Stratonovich transformation \cite{Guh06,KGG08} and the superbosonization formula \cite{Som07,LSZ07}. Thereby, we generalized both approaches. The superbosonization formula was proven in a new way and is now extended to odd dimensional supersymmetric Wishart--matrices in the Fermion--Fermion block for the quaternionic case. The generalized Hubbard--Stratonovich transformation was here extended to arbitrary dimensional supersymmetric Wishart--matrices which not only stem of averages over the matrix Green functions. \cite{GMW98,Zir06,Guh06,KKG08} Furthermore, we got an integral identity beyond the restriction of the matrix dimension, see Eq.~\eref{t1.0}. This approach distinguishes from the method presented in Ref.~\cite{BEKYZ07} by the integration of an additional matrix. It is, also, applicable on the artificial example $\beta=4$ and odd $b$ which has not been considered in Ref.~\cite{BEKYZ07}.

The generalized Hubbard--Stratonovich transformation and the superbosonization formula reduce in the absence of Grassmann variables to the ordinary integral identity for ordinary Wishart--matrices. \cite{Fyo02,LSZ07} In the general case with the restriction \eref{t1.0}, both approaches differ in the Fermion--Fermion block integration. Due to the Dirac--distribution and the differential operator, the integration over the non-compact domain in the generalized Hubbard--Stratonovich transformation is equal with help of the residue theorem to a contour integral. This contour integral is equivalent to the integration over the compact domain in the superbosonization formula. Hence, we found an integral identity between a compact integral and a differentiated Dirac--distribution.

\section*{Acknowledgements}

We thank Heiner Kohler for fruitful discussions. This work was supported by Deutsche Forschungsgemeinschaft within Sonderforschungsbereich Transregio 12 ``Symmetries and Universality in Mesoscopic Systems''.

\appendix

\section{ Proof of Theorem \ref{t1} (Superbosonization formula)}\label{app1}

First, we consider two particular cases. Let $d=0$ and $a\geq c$ be an arbitrary positive integer. Then, we find
\begin{equation}\label{p1.1}
 \widehat{B}\in\Sigma_{\beta,c0}^0=\Sigma_{\beta,c0}^{0(\dagger)}=\Sigma_{\beta,c0}^{0({\rm c})}\subset\Herm(\beta,c) .
\end{equation}
We introduce a Fourier--transformation
\begin{eqnarray}
   \fl\int\limits_{\mathbb{R}^{\beta ac}} F(\widehat{B})\exp\left(-\varepsilon\tr \widehat{B}\right)d[\widehat{V}]=\nonumber\\
 \fl= \displaystyle\left(\frac{\gamma_2}{2\pi}\right)^c\left(\frac{\gamma_2}{\pi}\right)^{\beta c(c-1)/2}\int\limits_{\Herm(\beta,c)}\int\limits_{\mathbb{R}^{\beta ac}} \mathcal{F}F(\sigma_1)\exp\left(\imath\tr \widehat{B}\sigma_1^+\right)d[\widehat{V}]d[\sigma_1]\label{p1.2}
\end{eqnarray}
where the measure $d[\sigma_1]$ is defined as in Eq.~\eref{t1.2} and $\sigma_1^+=\sigma_1+\imath\varepsilon \eins_{\gamma_2c}$. The Fourier--transform is
\begin{equation}\label{p1.3}
 \displaystyle\mathcal{F}F(\sigma_1)=\int\limits_{\Herm(\beta,c)}F(\rho_1)\exp\left(-\imath\tr\rho_1\sigma_1\right)d[\rho_1] .
\end{equation}
The integration over the supervectors, which are in this particular case ordinary vectors, yields
\begin{equation}\label{p1.4}
 \displaystyle\int\limits_{\mathbb{R}^{\beta ac}}\exp\left(\imath\tr\widehat{B}\sigma_1^+\right)d[\widehat{V}]=\det\left(\frac{\sigma_1^+}{\imath\gamma_1\pi}\right)^{-a/\gamma_1} .
\end{equation}
The Fourier--transform of this determinant is an Ingham--Siegel integral \cite{Ing33,Sie35}
\begin{equation}\label{p1.5}
  \fl\displaystyle\int\limits_{\Herm(\beta,c)}\exp\left(-\imath\tr \rho_1\sigma_1\right){\det}\left(-\imath\sigma_1^+\right)^{-a/\gamma_1}d[\sigma_1] = G_{a-c,c}^{(\beta)} \displaystyle\det \rho_1^{\kappa}\exp\left(-\varepsilon\tr \rho_1\right)\Theta(\rho_1),
\end{equation}
where the constant is
\begin{equation}\label{p1.6}
  G_{a-c,c}^{(\beta)}=  \left(\frac{\gamma_2}{\pi}\right)^{\gamma_2c\kappa}\prod\limits_{j=a-c+1}^a\frac{2\pi^{\beta j/2}}{\Gamma(\beta j/2)}
\end{equation}
and the exponent is
\begin{equation}\label{p1.7}
 \kappa=\frac{a-c+1}{\gamma_1}-\frac{1}{\gamma_2} .
\end{equation}
$\Gamma(.)$ is Euler's gamma--function. This integral was recently used in random matrix theory \cite{Fyo02} and is normalized in our notation as in Ref.~\cite{KGG08}. Thus, we find for Eq.~\eref{p1.2}
\begin{equation}\label{p1.8}
   \fl\int\limits_{\mathbb{R}^{\beta ac}} F(\widehat{B})\exp\left(-\varepsilon\tr \widehat{B}\right)d[\widehat{V}]=C_{ac0}^{(\beta)}\int\limits_{\Sigma_{\beta,c0}^{0({\rm c})}}F(\rho)\exp\left(-\varepsilon\tr\rho_1\right)\det\rho_1^{\kappa}d[\rho_1],
\end{equation}
which verifies this theorem. The product in the constant
\begin{equation}\label{p1.8b}
 C_{ac0}^{(\beta)}=2^{-c}\tilde{\gamma}^{\beta ac/2}\frac{{\rm Vol}\left(\U^{(\beta)}(a)\right)}{{\rm Vol}\left(\U^{(\beta)}(a-c)\right)}
\end{equation}
is a ratio of group volumes.

In the next case, we consider $c=0$ and arbitrary $d$. We see that
\begin{equation}\label{p1.10}
 \widehat{B}\in\Sigma_{\beta,0d}^{(\dagger)}
\end{equation}
is true. We integrate over
\begin{equation}\label{p1.11}
 \int\limits_{\Lambda_{2ad}} F(\widehat{B})\exp\left(\varepsilon\tr \widehat{B}\right)d[\widehat{V}],
\end{equation}
where the function $F$ is analytic. As in Ref.~\cite{Som07}, we expand $F(\widehat{B})\exp\left(\varepsilon\tr \widehat{B}\right)$ in the entries of $\widehat{B}$ and, then, integrate over every single term of this expansion. Every term is a product of $\widehat{B}$'s entries and can be generated by differentiation of $\left(\tr A\widehat{B}\right)^n$ with respect to $A\in\Sigma_{\beta,0d}^{0(\dagger)}$ for certain $n\in\mathbb{N}$. Thus, it is sufficient to proof the integral theorem
 \begin{equation}\label{p1.12}
  \int\limits_{\Lambda_{2ad}} \left(\tr A\widehat{B}\right)^nd[\widehat{V}]=C_{a0d}^{(\beta)}\int\limits_{\Sigma_{\beta,0d}^{0({\rm c})}}(\tr A\rho_2)^n\det \rho_2^{-\kappa}d[\rho_2] .
 \end{equation}
Since $\Sigma_{\beta,0d}^{0(\dagger)}$ is generated of $\Sigma_{\beta,0d}^{0({\rm c})}$ by analytic continuation in the eigenvalues, it is convenient that $A\in\Sigma_{\beta,0d}^{0({\rm c})}$. Then, $A^{-1/2}$ is well-defined and $A^{-1/2}\rho_2 A^{-1/2}\in\Sigma_{\beta,0d}^{0({\rm c})}$. We transform in Eq.~\eref{p1.13}
\begin{equation}\label{p1.13}
 \widehat{V}\rightarrow A^{-1/2}\widehat{V}\ ,\ \ \widehat{V}^{\dagger}\rightarrow \widehat{V}^\dagger A^{-1/2}\ \ {\rm and}\ \ \rho_2\rightarrow A^{-1/2}\rho_2 A^{-1/2} .
\end{equation}
The measures turns under this change into
\begin{eqnarray}
  d[\widehat{V}] & \rightarrow & \det A^{a/\gamma_1}d[\widehat{V}]\ \ {\rm and}\label{p1.14a}\\
  d[\rho_2] & \rightarrow & \det A^{-\kappa+a/\gamma_1}d[\rho_2]\label{p1.14b},
\end{eqnarray}
where the exponent is
\begin{equation}\label{p1.15}
 \kappa=\frac{a+1}{\gamma_1}+\frac{d-1}{\gamma_2} .
\end{equation}
Hence, we have to calculate the remaining constant defined by
 \begin{equation}\label{p1.16}
  \int\limits_{\Lambda_{2ad}} \left(\tr \widehat{B}\right)^nd[\widehat{V}]=C_{a0d}^{(\beta)}\int\limits_{\Sigma_{\beta,0d}^{0({\rm c})}}(\tr \rho_2)^n\det \rho_2^{-\kappa}d[\rho_2] .
 \end{equation}
This equation holds for arbitrary $n$. Then, this must also be valid for $F(\widehat{B})=\varepsilon=1$ in Eq.~\eref{p1.11}. The right hand side of Eq.~\eref{p1.11} is
\begin{equation}\label{p1.17}
 \int\limits_{\Lambda_{2ad}} \exp\left(\tr \widehat{B}\right)d[\widehat{V}]=\left(-2\pi\right)^{-ad} .
\end{equation}
On the left hand side, we first integrate over the group $\U^{(4/\beta)}\left(d\right)$ and get
\begin{eqnarray}
 \fl\displaystyle\int\limits_{\Sigma_{\beta,0d}^{0({\rm c})}}\exp\left(\tr\rho_2\right)\det \rho_2^{-\kappa}d[\rho_2]=\\
 \fl={\rm FU}_{d}^{(4/\beta)}\int\limits_{[0,2\pi]^{d}}|\Delta_{d}(e^{\imath\varphi_j})|^{4/\beta}\prod_{n=1}^d{\rm exp}\left(\gamma_1e^{\imath\varphi_n}\right)e^{-\imath\varphi_n(\gamma_1\kappa-1)}\frac{d\varphi_n}{2\pi} .\label{p1.18}
\end{eqnarray}
We derive this integral with help of Selberg's integral formula \cite{Meh04}. We assume that $\tilde{\beta}=4/\beta$  and $\gamma_1\kappa$ are arbitrary positive integers and $\tilde{\beta}$ is even. Then, we omit the absolute value and Eq.~\eref{p1.18} becomes
\begin{equation}\label{p1.19}
 \fl\displaystyle\int\limits_{\Sigma_{\beta,0d}^{0({\rm c})}}\exp\left(\tr\rho_2\right)\det \rho_2^{-\kappa}d[\rho_2]=\left.{\rm FU}_{d}^{(\beta)}\Delta_d^{\tilde{\beta}}\left(\frac{1}{\gamma_1}\frac{\partial}{\partial\lambda_j}\right)\prod_{n=1}^d\frac{\left(\gamma_1\lambda_n\right)^{\gamma_1\kappa-1}}{\Gamma\left(\gamma_1\kappa\right)}\right|_{\lambda=1} .
\end{equation}
We consider another integral which is the Laguerre version of Selberg's integral \cite{Meh04}
\begin{eqnarray}
  \fl\displaystyle\int\limits_{\mathbb{R}_+^{d}}\Delta_d^{\tilde{\beta}}(x)\prod\limits_{n=1}^d\exp\left(-\gamma_1x_n\right)x_n^{\xi}dx_n & = & \left.\Delta_d^{\tilde{\beta}}\left(-\frac{1}{\gamma_1}\frac{\partial}{\partial\lambda_j}\right)\prod_{n=1}^d\Gamma(\xi+1)\left(\gamma_1\lambda_n\right)^{-\xi-1}\right|_{\lambda=1}=\nonumber\\
  & = & \prod\limits_{n=1}^d\frac{\Gamma\left(1+n\tilde{\beta}/2\right)\Gamma\left(\xi+1+(n-1)\tilde{\beta}/2\right)}{\gamma_1^{\xi+1+\tilde{\beta}(d-1)/2}\Gamma\left(1+\tilde{\beta}/2\right)},\label{p1.20}
\end{eqnarray}
where $\xi$ is an arbitrary positive integer. Since $\tilde{\beta}$ is even the minus sign in the Vandermonde determinant vanishes. The equations \eref{p1.19} and \eref{p1.20} are up to the Gamma--functions polynomials in $\kappa$ and $\xi$. We remind that \eref{p1.20} is true for every complex $\xi$. Let $\RE\xi>0$, we have
\begin{eqnarray}
  \fl\left|\left.\Delta_d^{\tilde{\beta}}\left(-\frac{1}{\gamma_1}\frac{\partial}{\partial\lambda_j}\right)\prod_{n=1}^d\frac{\left(\gamma_1\lambda_n\right)^{-\xi-1}}{\Gamma\left(\xi+1+(n-1)\tilde{\beta}/2\right)}\right|_{\lambda=1}\right| & \leq & {\rm const.}<\infty\ \ {\rm and}\label{p1.21a}\\
  \fl\left|\gamma_1^{-d(\xi+1+\tilde{\beta}(d-1)/2)}\prod\limits_{n=1}^d\frac{\Gamma\left(1+n\tilde{\beta}/2\right)}{\Gamma(\xi+1)\Gamma\left(1+\tilde{\beta}/2\right)}\right| & \leq & {\rm const.}<\infty .\label{p1.21b}
\end{eqnarray}
The functions are bounded and regular for $\RE\xi>0$ and we can apply Carlson's theorem \cite{Meh04}. We identify $\xi=-\gamma_1\kappa$ and find
\begin{eqnarray}
 \fl\displaystyle\int\limits_{\Sigma_{\beta,0d}^{0({\rm c})}}\exp\left(\tr\rho_2\right)\det \rho_2^{-\kappa}d[\rho_2]=\nonumber\\
 \fl=\gamma_1^{ad}{\rm FU}_{d}^{(4/\beta)} \prod\limits_{n=1}^d\frac{\Gamma\left(1+n\tilde{\beta}/2\right)\Gamma\left(1-\gamma_1\kappa+(n-1)\tilde{\beta}/2\right)}{\Gamma\left(1+\tilde{\beta}/2\right)\Gamma\left(\gamma_1\kappa\right)\Gamma\left(1-\gamma_1\kappa\right)} .\label{p1.22}
\end{eqnarray}
Due to Euler's reflection formula $\Gamma(z)\Gamma(1-z)=\pi/\sin(\pi z)$, this equation simplifies to
\begin{equation}\label{p1.23}
 \fl\displaystyle\int\limits_{\Sigma_{\beta,0d}^{0({\rm c})}}\exp\left(\tr\rho_2\right)\det \rho_2^{-\kappa}d[\rho_2]=\gamma_1^{ad}{\rm FU}_{d}^{(4/\beta)} \prod\limits_{n=1}^d\frac{\imath^{4(n-1)/\beta}\Gamma\left(1+2n/\beta\right)}{\Gamma\left(1+2/\beta\right)\Gamma\left(\gamma_1\kappa-2(n-1)/\beta\right)}
\end{equation}
or equivalent
\begin{eqnarray}
  \fl\displaystyle2^{\tilde{\beta}d(d-1)/2}\int\limits_{[0,2\pi]^{d}}\prod\limits_{1\leq n<m\leq d}\left|\sin\left(\frac{\varphi_n-\varphi_m}{2}\right)\right|^{\tilde{\beta}}\prod_{n=1}^d{\rm exp}\left(\gamma_1e^{\imath\varphi_n}\right)e^{-\imath\varphi_na}\frac{d\varphi_n}{2\pi}=\nonumber\\
  \fl= \displaystyle\gamma_1^{ad}\prod\limits_{n=1}^d\frac{\Gamma\left(1+n\tilde{\beta}/2\right)}{\Gamma\left(1+\tilde{\beta}/2\right)\Gamma\left(a+1+(n-1)\tilde{\beta}/2\right)} .\label{p1.24}
\end{eqnarray}
Since $a$ is a positive integer for all positive and even $\tilde{\beta}$, the equations above are true for all such $\tilde{\beta}$. For constant natural numbers $a$, $d$, $\gamma_1$ and complex $\tilde{\beta}$ with $\RE\tilde{\beta}>0$, the inequalities
\begin{eqnarray}
 \fl\displaystyle\left|\int\limits_{[0,2\pi]^{d}}\prod\limits_{1\leq n<m\leq d}\left|\sin\left(\frac{\varphi_n-\varphi_m}{2}\right)\right|^{\tilde{\beta}}\prod_{n=1}^d{\rm exp}\left(\gamma_1e^{\imath\varphi_n}\right)e^{-\imath\varphi_na}\frac{d\varphi_n}{2\pi}\right|\leq\nonumber\\
 \fl\leq \displaystyle\int\limits_{[0,2\pi]^{d}}\prod\limits_{1\leq n<m\leq d}\left|\sin\left(\frac{\varphi_n-\varphi_m}{2}\right)\right|^{\RE\tilde{\beta}}\prod_{n=1}^d{\rm exp}\left(\gamma_1\cos\varphi_n\right)\frac{d\varphi_n}{2\pi}<\infty\ \ {\rm and}\label{p1.25a}\\
 \fl \displaystyle\left|2^{-\tilde{\beta}d(d-1)/2}\gamma_1^{ad}\prod\limits_{n=1}^d\frac{\Gamma\left(1+n\tilde{\beta}/2\right)}{\Gamma\left(1+\tilde{\beta}/2\right)\Gamma\left(a+1+(n-1)\tilde{\beta}/2\right)}\right|\leq\nonumber\\
 \fl\leq \displaystyle{\rm const.}\ 2^{-\RE\tilde{\beta}d(d-1)/2}<\infty\label{p1.25b}
\end{eqnarray}
are valid and allow us with Carlson's theorem to extend Eq.~\eref{p1.24} to every complex $\tilde{\beta}$, in particular to $\tilde{\beta}=1$. Thus, we find for the constant in Eq.~\eref{p1.16}
\begin{equation}\label{p1.26}
  C_{a0d}= \left(-2\pi\gamma_1\right)^{-ad}\left[\prod\limits_{n=1}^d\frac{\imath^{4(n-1)/\beta}\pi^{2(n-1)/\beta}}{\Gamma(a+1+2(n-1)/\beta)}\right]^{-1} .
\end{equation}

Now, we consider arbitrary $d$ and $a\geq c$ and split
\begin{equation}\label{p1.27}
 \Psi_{j1}^{({\rm C})}=\left[\begin{array}{c} \mathbf{x}_j \\ \chi_j \end{array}\right]
\end{equation}
and
\begin{equation}\label{p1.28}
 \widehat{B}=\frac{1}{\tilde{\gamma}}\sum\limits_{j=1}^a\Psi_{j1}^{({\rm C})}\Psi_{j1}^{({\rm C})\dagger}=\left[\begin{array}{c|c} \displaystyle\sum\limits_{j=1}^a\frac{\mathbf{x}_j\mathbf{x}_j^\dagger}{\tilde{\gamma}} & \displaystyle\sum\limits_{j=1}^a\frac{\mathbf{x}_j\chi_j^\dagger}{\tilde{\gamma}} \\ \hline \displaystyle\sum\limits_{j=1}^a\frac{\chi_j\mathbf{x}_j^\dagger}{\tilde{\gamma}} & \displaystyle\sum\limits_{j=1}^a\frac{\chi_j\chi_j^\dagger}{\tilde{\gamma}} \end{array}\right]=\left[\begin{array}{cc} B_{11} & B_{12} \\ B_{21} & B_{22} \end{array}\right]
\end{equation}
such that $\mathbf{x}_j$ contains all commuting variables of $\Psi_{j1}^{({\rm C})}$ and $\chi_j$ depends on all Grassmann variables. Then, we replace the sub-matrices $B_{12},B_{21}$ and $B_{22}$ by Dirac--distributions
\begin{eqnarray}
  \fl\int\limits_{\mathfrak{R}} F(\widehat{B})\exp\left(-\varepsilon\Str \widehat{B}\right)d[\widehat{V}]=\nonumber\\
  \fl= C_1\int\limits_{\Herm(4/\beta,d)^2}\int\limits_{\mathfrak{R}}\int\limits_{\left(\Lambda_{2cd}\right)^2}d[\eta]d[\tilde{\eta}]d[\widehat{V}]d[\tilde{\rho}_2]d[\sigma_2] F\left(\left[\begin{array}{cc} B_{11} & \rho_\eta \\ -\rho_\eta^\dagger & \tilde{\rho_2} \end{array}\right]\right)\times\nonumber\\
  \fl\times {\rm exp}\left[-\varepsilon\Str B-\imath\left(\tr(\rho_\eta^\dagger+B_{21})\sigma_{\tilde{\eta}}+\tr\sigma_{\tilde{\eta}}^\dagger(\rho_\eta-B_{12})-\tr(\tilde{\rho_2}-B_{22})\sigma_2\right)\right],\label{p1.29}
\end{eqnarray}
where
\begin{equation}\label{p1.30}
 C_1=\left(\frac{2\pi}{\tilde{\gamma}}\right)^{2cd}\left(\frac{\gamma_1}{\pi}\right)^{2d(d-1)/\beta}\left(\frac{\gamma_1}{2\pi}\right)^d .
\end{equation}
The matrices $\rho_\eta$ and $\sigma_{\tilde{\eta}}$ are rectangular matrices depending on Grassmann variables as in the Boson--Fermion and Fermion--Boson block in the sets (\ref{4.1}-\ref{4.3}). Shifting $\chi_j\rightarrow\chi_j+\left(\sigma_2^+\right)^{-1}\sigma_{\tilde{\eta}}^\dagger\mathbf{x}_j$ and $\chi_j^\dagger\rightarrow\chi_j^\dagger-\mathbf{x}_j^\dagger\sigma_{\tilde{\eta}}\left(\sigma_2^+\right)^{-1}$, we get
\begin{eqnarray}
  \fl\int\limits_{\mathfrak{R}} F(\widehat{B})\exp\left(-\varepsilon\Str \widehat{B}\right)d[\widehat{V}]=\nonumber\\
  \fl=  C_1\int\limits_{\Herm(4/\beta,d)^2}\int\limits_{\mathfrak{R}}\int\limits_{\left(\Lambda_{2cd}\right)^2}d[\eta]d[\tilde{\eta}]d[\widehat{V}]d[\tilde{\rho}_2]d[\sigma_2] F\left(\left[\begin{array}{cc} B_{11} & \rho_\eta \\ -\rho_\eta^\dagger & \tilde{\rho_2} \end{array}\right]\right)\times\nonumber\\
  \fl\times {\rm exp}\left[-\varepsilon\Str B-\imath\left(\tr\sigma_{\tilde{\eta}}^\dagger B_{11}\sigma_{\tilde{\eta}}\left(\sigma_2^+\right)^{-1}+\tr\rho_\eta^\dagger\sigma_{\tilde{\eta}}+\tr\sigma_{\tilde{\eta}}^\dagger\rho_\eta-\tr(\tilde{\rho_2}-B_{22})\sigma_2\right)\right].\label{p1.31}
\end{eqnarray}
This integral only depends on $B_{11}$ and $B_{22}$. Thus, we apply the first case of this proof and replace $B_{11}$. We find
\begin{eqnarray}
  \fl\int\limits_{\mathfrak{R}}F(\widehat{B})\exp\left(-\varepsilon\Str \widehat{B}\right)d[\widehat{V}]=\nonumber\\
  \fl=  C_{ac0}^{(\beta)}C_1\int\limits_{\Herm(4/\beta,d)^2}\int\limits_{\mathfrak{R}}\int\limits_{\left(\Lambda_{2cd}\right)^2} d[\chi]d[\eta]d[\tilde{\eta}]d[\rho_1]d[\tilde{\rho}_2]d[\sigma_2]F\left(\left[\begin{array}{cc} B_{11} & \rho_\eta \\ -\rho_\eta^\dagger & \tilde{\rho_2} \end{array}\right]\right)\det\rho_1^{\tilde{\kappa}}\times\nonumber\\
  \fl\times \displaystyle {\rm exp}\left[\varepsilon(\tr B_{22}-\tr\rho_1)+\imath\left(\tr\sigma_{\tilde{\eta}}^\dagger\rho_1\sigma_{\tilde{\eta}}\left(\sigma_2^+\right)^{-1}-\tr\rho_\eta^\dagger\sigma_{\tilde{\eta}}-\tr\sigma_{\tilde{\eta}}^\dagger\rho_\eta+\tr(\tilde{\rho_2}-B_{22})\sigma_2\right)\right] \label{p1.32}
\end{eqnarray}
with the exponent
\begin{equation}\label{p1.33}
 \tilde{\kappa}=\frac{a-c+1}{\gamma_1}-\frac{1}{\gamma_2} .
\end{equation}
After another shifting $\sigma_{\tilde{\eta}}\rightarrow\sigma_{\tilde{\eta}}-\rho_1^{-1}\rho_\eta\sigma_2^+$ and $\sigma_{\tilde{\eta}}^\dagger\rightarrow\sigma_{\tilde{\eta}}^\dagger-\sigma_2^+\rho_\eta^\dagger\rho_1^{-1}$, we integrate over $d[\tilde{\eta}]$ and $B_{22}$ and have
\begin{eqnarray}
  \fl\int\limits_{\mathfrak{R}}F(\widehat{B})\exp\left(-\varepsilon\Str \widehat{B}\right)d[\widehat{V}]=\nonumber\\
  \fl= C_{ac0}^{(\beta)}C_2\int\limits_{\Sigma_{\beta,c0}^{0({\rm c})}}\int\limits_{\Herm(4/\beta,d)^2}\int\limits_{\Lambda_{2cd}}d[\eta]d[\rho_1]d[\tilde{\rho}_2]d[\sigma_2] F\left(\left[\begin{array}{cc} \rho_1 & \rho_\eta \\ -\rho_\eta^\dagger & \tilde{\rho_2} \end{array}\right]\right)\times\nonumber\\
  \fl\times \displaystyle\det\rho_1^{\kappa}\det\left(\sigma_2^+\right)^{(a-c)/\gamma_1}{\rm exp}\left[-\varepsilon\tr\rho_1+\imath\left(\tr\rho_\eta^\dagger\rho_1^{-1}\rho_\eta\sigma_2^++\tr\tilde{\rho_2}\sigma_2\right)\right],\label{p1.34}
\end{eqnarray}
where the exponent is
\begin{equation}\label{p1.35}
 \kappa=\frac{a-c+1}{\gamma_1}+\frac{d-1}{\gamma_2}
\end{equation}
and the new constant is
\begin{equation}\label{p1.36}
 C_2=\left(\frac{\imath}{2\pi}\right)^{ad}\left(\frac{2\pi}{\tilde{\gamma}\imath}\right)^{cd}\left(\frac{\gamma_1}{\pi}\right)^{2d(d-1)/\beta}\left(\frac{\gamma_1}{2\pi}\right)^d .
\end{equation}
We express the determinant in $\sigma_2^+$ as in Sec. \ref{sec2} as Gaussian integrals and define a new $(\gamma_2(a-c)+0)\times(0+\gamma_1d)$ rectangular supermatrix $\widehat{V}_{{\rm new}}$ and its corresponding $(0+\gamma_1d)\times(0+\gamma_1d)$ supermatrix $\widehat{B}_{{\rm new}}=\tilde{\gamma}^{-1}\widehat{V}_{{\rm new}}\widehat{V}_{{\rm new}}^\dagger$. Integrating $\sigma_2$ and $\rho_2$, Eq.~\eref{p1.34} becomes
\begin{eqnarray}
  \fl\int\limits_{\mathfrak{R}} F(\widehat{B})\exp\left(-\varepsilon\Str \widehat{B}\right)d[\widehat{V}]=\tilde{\gamma}^{-cd}C_{ac0}^{(\beta)}\int\limits_{\Sigma_{\beta,c0}^{0({\rm c})}}\int\limits_{\Lambda_{2(a-c)d}} F\left(\left[\begin{array}{c|c} \rho_1 & \rho_\eta \\ \hline -\rho_\eta^\dagger & \widehat{B}_{{\rm new}}-\rho_\eta^\dagger\rho_1^{-1}\rho_\eta \end{array}\right]\right)\times\nonumber\\
  \fl\times \displaystyle \exp\left(-\varepsilon\tr\rho_1+\varepsilon\tr (\widehat{B}_{{\rm new}}-\eta^\dagger\rho_1^{-1}\eta)\right) \det\rho_1^{\kappa}d[\widehat{V}_{{\rm new}}]d[\eta]d[\rho_1] .\label{p1.37}
\end{eqnarray}
Now, we apply the second case in this proof and shift $\rho_2\rightarrow\rho_2+\rho_\eta^\dagger\rho_1^{-1}\rho_\eta$ by analytic continuation. We get the final result
\begin{equation}
 \fl\int\limits_{\mathfrak{R}} F(\widehat{B})\exp\left(-\varepsilon\Str \widehat{B}\right)d[\widehat{V}]=C_{acd}^{(\beta)}\int\limits_{\Sigma_{\beta,cd}^{0({\rm c})}} F\left(\rho\right)\displaystyle \exp\left(-\varepsilon\Str\rho\right) \Sdet\rho^{\kappa}d[\rho]\label{p1.39}
\end{equation}
with
\begin{eqnarray}
 \fl C_{acd}^{(\beta)} = \tilde{\gamma}^{-cd}C_{ac0}^{(\beta)}C_{a-c,0d}^{(\beta)}=\nonumber\\
 \fl=\left(-2\pi\gamma_1\right)^{-ad}\left(-\frac{2\pi}{\gamma_2}\right)^{cd}2^{-c}\tilde{\gamma}^{\beta ac/2}\frac{{\rm Vol}\left(\U^{(\beta)}(a)\right)}{{\rm Vol}\left(\U^{(\beta)}(a-c)\right)}\prod\limits_{n=1}^d\frac{\Gamma\left(\gamma_1\kappa+2(n-d)/\beta\right)}{\imath^{4(n-1)/\beta}\pi^{2(n-1)/\beta}}=\nonumber\\
 \fl= \imath^{-2d(d-1)/\beta}\frac{(2\pi)^d\tilde{\gamma}^{\beta ac/2-cd}}{(-2)^{(c-a)d}2^{c}}
  \left\{\begin{array}{ll}
   \displaystyle\frac{2^{d^2}{\rm Vol}\left(\U^{(1)}(a)\right)}{{\rm Vol}\left(\U^{(1)}(a-c+2d)\right)} & ,\ \beta=1\\
   \displaystyle\frac{{\rm Vol}(\U^{(2)}(a)}{{\rm Vol}\left(\U^{(2)}(a-c+d)\right)} & ,\ \beta=2\\
   \displaystyle\frac{2^{-(2a+1-c)c}{\rm Vol}\left(\U^{(1)}(2a+1)\right)}{{\rm Vol}\left(\U^{(1)}(2(a-c)+d+1)\right)} & ,\ \beta=4
  \end{array}\right.\ .\label{p1.40}
\end{eqnarray}

\section{ Proof of Theorem \ref{t2} (Generalized Hubbard--Stratonovich transformation)}\label{app2}

We choose a Wick--rotation $e^{\imath\psi}$ that all calculations below are well defined. Then, we perform a Fourier transformation
\begin{equation}\label{p2.1}
 \fl\displaystyle\int\limits_{\mathfrak{R}}F(\widehat{B})\exp\left(-\varepsilon\Str \widehat{B}\right)d[\widehat{V}]=\widetilde{C}_1\int\limits_{\widetilde{\Sigma}_{\beta,cd}^{(-\psi)}}\int\limits_{\mathfrak{R}}\mathcal{F}F(\sigma)\exp\left(\imath\Str \widehat{B}\sigma^+\right)d[\widehat{V}]d[\sigma],
\end{equation}
where $\sigma^+=\sigma+\imath\varepsilon \eins_{\gamma_2c+\gamma_1d}$,
\begin{equation}\label{p2.2}
 \mathcal{F}F(\sigma)=\int\limits_{\widetilde{\Sigma}_{\beta,cd}^{(\psi)}}F(\rho)\exp\left(-\imath\Str\rho\sigma\right)d[\rho],
\end{equation}
and the constant is
\begin{equation}\label{p2.3}
 \displaystyle\widetilde{C}_1=\left(\frac{2\pi}{\tilde{\gamma}}\right)^{2cd}\left(\frac{\gamma_2}{2\pi}\right)^{c}\left(\frac{\gamma_2}{\pi}\right)^{\beta c(c-1)/2}\left(\frac{\gamma_1}{2\pi}\right)^{d}\left(\frac{\gamma_1}{\pi}\right)^{2d(d-1)/\beta} .
\end{equation}
The integration over $\widehat{V}$ yields
\begin{equation}\label{p2.4}
 \displaystyle\int\limits_{\mathfrak{R}}F(\widehat{B})\exp\left(-\varepsilon\Str \widehat{B}\right)d[\widehat{V}]=\widetilde{C}_2\int\limits_{\widetilde{\Sigma}_{\beta,cd}^{(-\psi)}}\mathcal{F}F(\sigma)\Sdet^{-a/\gamma_1}\sigma^+d[\sigma]
\end{equation}
with
\begin{equation}\label{p2.5}
 \fl\displaystyle\widetilde{C}_2=\left(\frac{2\pi}{\tilde{\gamma}}\right)^{2cd}\left(\frac{\gamma_2}{2\pi}\right)^{c}\left(\frac{\gamma_2}{\pi}\right)^{\beta c(c-1)/2}\left(\frac{\gamma_1}{2\pi}\right)^{d}\left(\frac{\gamma_1}{\pi}\right)^{2d(d-1)/\beta}\left(\frac{\imath}{2\pi}\right)^{ad}\left(\gamma_1\pi\imath\right)^{\beta ac/2} .
\end{equation}
We transform this result back by a Fourier--transformation
\begin{equation}\label{p2.6}
 \fl\displaystyle\int\limits_{\mathfrak{R}}F(\widehat{B})\exp\left(-\varepsilon\Str \widehat{B}\right)d[\widehat{V}]=\widetilde{C}_2\int\limits_{\widetilde{\Sigma}_{\beta,cd}^{(\psi)}}F(\rho)I_{cd}^{(\beta,a)}(\rho)\exp\left(-\varepsilon\Str\rho\right)d[\rho],
\end{equation}
where we have to calculate the supersymmetric Ingham--Siegel integral
\begin{equation}\label{p2.7}
 I_{cd}^{(\beta,a)}(\rho)=\int\limits_{\widetilde{\Sigma}_{\beta,cd}^{(-\psi)}}\exp\left(-\imath\Str\rho\sigma^+\right)\Sdet^{-a/\gamma_1}\sigma^+d[\sigma] .
\end{equation}
This distribution is rotation invariant under $\U^{(\beta)}(c/d)$. The ordinary version, $d=0$, of Eq.~\eref{p2.6} is Eq.~\eref{p1.5}.

After performing four shifts
\begin{eqnarray}
  \sigma_1 & \rightarrow & \sigma_1-\sigma_{\tilde{\eta}}\left(\sigma_2+\imath e^{\imath\psi}\varepsilon \eins_{\gamma_1d}\right)^{-1}\sigma_{\tilde{\eta}}^\dagger ,\label{p2.8a}\\
  \sigma_{\tilde{\eta}} & \rightarrow & \sigma_{\tilde{\eta}}-\rho_1^{-1}\rho_\eta\left(\sigma_2+\imath e^{\imath\psi}\varepsilon \eins_{\gamma_1d}\right) ,\label{p2.8b}\\
  \sigma_{\tilde{\eta}}^\dagger & \rightarrow & \sigma_{\tilde{\eta}}^\dagger-\left(\sigma_2+\imath e^{\imath\psi}\varepsilon \eins_{\gamma_1d}\right)\rho_\eta^\dagger\rho_1^{-1} ,\label{p2.8c}\\
  \rho_2 & \rightarrow & \rho_2-\rho_\eta^\dagger\rho_1^{-1}\rho_\eta ,\label{p2.8d}
\end{eqnarray}
and defining
 \begin{equation}\label{p2.8e}
  \hat{\rho}=\left[\begin{array}{c|c} \rho_1 & e^{\imath\psi/2}\rho_\eta \\ \hline -e^{\imath\psi/2}\rho_\eta^\dagger & e^{\imath\psi}\left(\rho_2-\rho_\eta^\dagger\rho_1^{-1}\rho_\eta\right) \end{array}\right],
 \end{equation}
we find
\begin{equation}\label{p2.9}
 \displaystyle\int\limits_{\mathfrak{R}}F(\widehat{B})\exp\left(-\varepsilon\Str \widehat{B}\right)d[\widehat{V}]=\widetilde{C}_2\int\limits_{\widetilde{\Sigma}_{\beta,cd}^{(\psi)}}F\left(\hat{\rho}\right)\widetilde{I}(\rho)\exp\left(-\varepsilon\Str\hat{\rho}\right)d[\rho],
\end{equation}
where
\begin{eqnarray}
  \fl\displaystyle\widetilde{I}(\rho) = \displaystyle\int\limits_{\widetilde{\Sigma}_{\beta,cd}^{(-\psi)}}{\rm exp}\left[{\varepsilon\Str\rho-\imath\left(\tr\rho_1\sigma_1-\tr\rho_2\sigma_2+\tr\sigma_{\tilde{\eta}}^\dagger\rho_1\sigma_{\tilde{\eta}}(\sigma_2+\imath e^{\imath\psi}\varepsilon \eins_{\gamma_1d})^{-1}\right)}\right]\times\nonumber\\
  \times \displaystyle\left(\frac{\det(e^{-\imath\psi}\sigma_2+\imath \varepsilon \eins_{\gamma_1d})}{\det(\sigma_1+\imath\varepsilon \eins_{\gamma_2c})}\right)^{a/\gamma_1}d[\sigma] .\label{p2.10}
\end{eqnarray}
We integrate over $d[\tilde{\eta}]$ and apply Eq.~\eref{p1.5} for the $d[\sigma_1]$--integration. Then, Eq.~\eref{p2.10} reads
\begin{eqnarray}
  \fl\displaystyle\widetilde{I}(\rho) = \displaystyle\widetilde{C}_3e^{-\imath\psi cd}\det\rho_1^\kappa\Theta(\rho_1)\times\nonumber\\
 \fl\times \int\limits_{\Herm(4/\beta,d)}\exp\left(-\imath\tr\rho_2(\sigma_2+\imath e^{\imath\psi}\varepsilon \eins_{\gamma_1d})\right)\det(e^{-\imath\psi}\sigma_2+\imath \varepsilon)^{(a-c)/\gamma_1}d[\sigma_2]\label{p2.11}
\end{eqnarray}
with the constant
\begin{equation}\label{p2.12}
  \displaystyle\widetilde{C}_3=\imath^{-\beta ac/2}\left(\frac{\tilde{\gamma}}{2\pi\imath}\right)^{cd}G_{a-c,c}^{(\beta)},
\end{equation}
see Eq.~\eref{p1.6}. The exponent $\kappa$ is the same as in Eq.~\eref{t1.3}. As in Ref.~\cite{KGG08}, we decompose $\sigma_2$ in angles and eigenvalues and integrate over the angles. Thus, we get the ordinary matrix Bessel--function
\begin{equation}\label{p2.13}
 \varphi_d^{(4/\beta)}(r_2,s_2)=\int\limits_{\U^{(4/\beta)}(d)}\exp\left(\imath\tr r_2Us_2U^\dagger\right) d\mu(U)
\end{equation}
in Eq.~\eref{p2.11} which are only for certain $\beta$ and $d$ explicitly known. However, the analog of the Sekiguchi differential operator for the ordinary matrix Bessel--functions $D_{dr_2}^{(4/\beta)}$, see Eq.~\eref{t2.2}, fulfills the eigenvalue equation
\begin{equation}\label{p2.14}
 D_{dr_2}^{(4/\beta)}\varphi_d^{(4/\beta)}(r_2,s_2)=(\imath\gamma_1)^d\det s_2^{1/\gamma_1}\varphi_d^{(4/\beta)}(r_2,s_2) .
\end{equation}
Since the determinant of $\sigma_2$ stands in the numerator, we shift $\sigma_2\rightarrow\sigma_2-\imath e^{\imath\psi}\varepsilon \eins_{\gamma_1d}$ and replace the determinants in Eq.~\eref{p2.11} by $D_{dr_2}^{(4/\beta)}$. After an integration over $\sigma_2$, we have
\begin{equation}\label{p2.15}
  \displaystyle\widetilde{I}(\rho)= \displaystyle\widetilde{C}_4e^{-\imath\psi cd}\det\rho_1^\kappa\Theta(\rho_1)\left(e^{-\imath\psi d}D_{dr_2}^{(4/\beta)}\right)^{a-c}\frac{\delta(r_2)}{|\Delta_d(e^{\imath\psi}r_2)|^{4/\beta}} .
\end{equation}
The constant is
\begin{equation}\label{p2.16}
 \fl\widetilde{C}_4=\imath^{-\beta ac/2}\left(\frac{\tilde{\gamma}}{2\pi\imath}\right)^{cd}G_{a-c,c}^{(\beta)}(\imath\gamma_1)^{(c-a)d}\left(\frac{\pi}{\gamma_1}\right)^{2d(d-1)/\beta}\left(\frac{2\pi}{\gamma_1}\right)^d\frac{1}{{\rm FU}_d^{(4/\beta)}} .
\end{equation}
Summarizing the constants \eref{p2.5} and \eref{p2.16}, we get
\begin{equation}\label{p2.17}
 \fl\widetilde{C}_{acd}^{(\beta)}=\widetilde{C}_2\widetilde{C}_4=2^{-c}\left(2\pi\gamma_1\right)^{-ad}\left(\frac{2\pi}{\gamma_2}\right)^{cd}\tilde{\gamma}^{\beta ac/2}\frac{{\rm Vol}\left(\U^{(\beta)}(a)\right)}{{\rm Vol}\left(\U^{(\beta)}(a-c)\right){\rm FU}_d^{(4/\beta)}} .
\end{equation}
Due to the Dirac--distribution, we shift $D_{dr_2}^{(4/\beta)}$ from the Dirac--distribution to the superfunction and remove the Wick--rotation. Hence, we find the result of the Theorem.

\section{ Proof of Theorem \ref{t3} (Equivalence of both approaches)}\label{app3}

We define the function
\begin{eqnarray}
  \widetilde{F}(r_2) & = & \int\limits_{\U^{4/\beta}(d)}\int\limits_{\Herm(\beta,c)}\int\limits_{\Lambda_{2cd}}F\left(\left[\begin{array}{c|c} \rho_1 & \rho_\eta \\ \hline -\rho_\eta^\dagger & Ur_2U^\dagger-\rho_\eta^\dagger\rho_1^{-1}\rho_\eta \end{array}\right]\right)\times\nonumber\\
  & \times & \exp\left[-\varepsilon(\tr\rho_1-\tr (r_2-\rho_\eta^\dagger\rho_1^{-1}\rho_\eta)\right]{\det}^\kappa\rho_1d[\eta]d[\rho_1]d\mu(U) .\label{p3.1}
\end{eqnarray}
Then, we have to prove
\begin{equation}\label{p3.2}
 \fl C_{acd}^{(\beta)}\int\limits_{[0,2\pi]^{d}}\widetilde{F}\left(e^{\imath\varphi_j}\right)\left|\Delta_d\left(e^{\imath\varphi_j}\right)\right|^{4/\beta}\prod\limits_{n=1}^d\frac{e^{\imath(1-\kappa)\varphi_n}d\varphi_n}{2\pi}=\widetilde{C}_{acd}^{(\beta)}\left.\left((-1)^dD_{dr_2}^{(4/\beta)}\right)^{a-c}\widetilde{F}(r_2)\right|_{r_2=0}\hspace*{-0.5cm}.
\end{equation}
Since $\widetilde{F}$ is permutation invariant and a Schwartz--function, we express $\widetilde{F}$ as an integral over ordinary matrix Bessel--functions,
\begin{equation}\label{p3.3}
 \widetilde{F}(r_2)=\int\limits_{\mathbb{R}^d}g(q)\varphi_d^{(4/\beta)}(r_2,q)|\Delta_d(q)|^{4/\beta}dq,
\end{equation}
where $g$ is a Schwartz--function. The integral and the differential operator in Eq.~\eref{p3.2} commute with the integral in Eq.~\eref{p3.3}. Thus, we only need to prove
\begin{eqnarray}
  & & C_{acd}^{(\beta)}\int\limits_{[0,2\pi]^{d}}\varphi_d^{(4/\beta)}\left(e^{\imath\varphi_j},q\right)\left|\Delta_d\left(e^{\imath\varphi_j}\right)\right|^{4/\beta}\prod\limits_{n=1}^d\frac{e^{\imath(1-\gamma_1\kappa)\varphi_n}d\varphi_n}{2\pi}=\nonumber\\
  & = & \widetilde{C}_{acd}^{(\beta)}\left.\left((-1)^dD_{dr_2}^{(4/\beta)}\right)^{a-c}\varphi_d^{(4/\beta)}(r_2,q)\right|_{r_2=0}\label{p3.4}
\end{eqnarray}
for all $q\in\mathcal{S}_1^d$ where $\mathcal{S}_1$ is the unit--circle in the complex plane. The right hand side of this equation is with help of Eq.~\eref{p2.14}
\begin{equation}\label{p3.5}
 \left.\left(D_{dr_2}^{(4/\beta)}\right)^{a-c}\varphi_d^{(4/\beta)}(r_2,q)\right|_{r_2=0}=(-\imath\gamma_1)^{d(a-c)}\det q^{(a-c)/\gamma_1} .
\end{equation}
The components of $q$ are complex phase factors. The integral representation of the ordinary matrix Bessel--functions \eref{p2.13} and the $d[\varphi]$--integral in Eq.~\eref{p3.4} form the integral over the circular ensembles $\CU^{(4/\beta)}(d)$. Thus, $q$ can be absorbed by $e^{\imath\varphi_j}$ and we find
\begin{eqnarray}
  & & \int\limits_{[0,2\pi]^{d}}\varphi_d^{(4/\beta)}\left(e^{\imath\varphi_j},q\right)\left|\Delta_d\left(e^{\imath\varphi_j}\right)\right|^{4/\beta}\prod\limits_{n=1}^d\frac{e^{\imath(1-\gamma_1\kappa)\varphi_n}d\varphi_n}{2\pi}=\nonumber\\
  & = & \det q^{(a-c)/\gamma_1}\int\limits_{[0,2\pi]^{d}}\varphi_d^{(4/\beta)}\left(e^{\imath\varphi_j},1\right)\left|\Delta_d\left(e^{\imath\varphi_j}\right)\right|^{4/\beta}\prod\limits_{n=1}^d\frac{e^{\imath(1-\gamma_1\kappa)\varphi_n}d\varphi_n}{2\pi} .\label{p3.6}
\end{eqnarray}
The ordinary matrix Bessel--function is at $q=1$ the exponential function
\begin{equation}\label{p3.7}
 \varphi_d^{(4/\beta)}\left(e^{\imath\varphi_j},1\right)={\rm exp}\left(\imath\gamma_1\sum\limits_{n=1}^de^{\imath\varphi_n}\right) .
\end{equation}
With Eq.~\eref{p1.24}, the integral on the left hand side in Eq.~\eref{p3.6} yields with this exponential function
\begin{eqnarray}
   & & \int\limits_{[0,2\pi]^{d}}\left|\Delta_d\left(e^{\imath\varphi_j}\right)\right|^{4/\beta}\prod\limits_{n=1}^d\frac{e^{\imath(1-\gamma_1\kappa)\varphi_n}{\rm exp}\left(\imath\gamma_1e^{\imath\varphi_n}\right)d\varphi_n}{2\pi}=\nonumber\\
  & = & (\imath\gamma_1)^{(a-c)d}\prod\limits_{n=1}^d\frac{\imath^{4(n-1)/\beta}\Gamma\left(1+2n/\beta\right)}{\Gamma\left(1+2/\beta\right)\Gamma\left(a-c+1+2(n-1)/\beta\right)}=\nonumber\\
  & = & \frac{(\imath\gamma_1)^{(a-c)d}}{{\rm FU}_d^{(4/\beta)}}\prod\limits_{n=1}^d\frac{\imath^{4(n-1)/\beta}\pi^{2(n-1)/\beta}}{\Gamma\left(a-c+1+2(n-1)/\beta\right)} .\label{p3.8}
\end{eqnarray}
Hence, the normalization on both sides in Eq.~\eref{p3.2} is equal.

\section*{References}

\end{document}